\documentclass[a4paper,fleqn]{cas-sc}

\usepackage[utf8]{inputenc}
\usepackage[english]{babel}

\usepackage[numbers]{natbib}
\usepackage{amssymb}
\usepackage{gensymb}
\usepackage{graphicx}
\usepackage{subcaption}
\usepackage{afterpage}
\usepackage{mhchem}
\usepackage{chemformula}

\usepackage{framed} % Framing content
\usepackage{multicol} % Multiple columns environment
\usepackage{nomencl} % Nomenclature package
\makenomenclature
\renewcommand*\nompreamble{\begin{multicols}{2}}
\renewcommand*\nompostamble{\end{multicols}}

\usepackage{etoolbox}
\renewcommand\nomgroup[1]{%
  \item[\bfseries
  \ifstrequal{#1}{A}{Alphabetic}{%
  \ifstrequal{#1}{S}{Subscripts}{%
  \ifstrequal{#1}{O}{Other Symbols}{
   \ifstrequal{#1}{G}{Greek}{%
   \ifstrequal{#1}{B}{Abbreviations}{}}}}}
]}

\def\tsc#1{\csdef{#1}{\textsc{\lowercase{#1}}\xspace}}
\tsc{WGM}
\tsc{QE}
\tsc{EP}
\tsc{PMS}
\tsc{BEC}
\tsc{DE}
\begin{document}
\let\WriteBookmarks\relax
\def\floatpagepagefraction{1}
\def\textpagefraction{.001}
\shorttitle{Tetralin+C$_{60}$}
%\shortauthors{Rita Adrião Lamosa et~al.}
%\begin{frontmatter}

\title [mode = title]{Tetralin + fullerene C$_{60}$ solutions for thermal management of flat-plate photovoltaic/thermal collector}             
%\tnotemark[1,2]

%\tnotetext[1]{This document is the results of the research
%   project funded by the National Science Foundation.}

%\tnotetext[2]{The second title footnote which is a longer text matter
%   to fill through the whole text width and overflow into
%   another line in the footnotes area of the first page.}

\author[1]{Rita Adrião Lamosa}
\author[2]{Igor Motovoy}[orcid=0000-0002-1409-4453]
\author[3]{Nikita Khliiev}
\author[4]{Artem Nikulin}[orcid=0000-0002-3304-0506]\cormark[1]
\author[2,5]{Olga Khliyeva}[orcid=0000-0002-3592-4989]\cormark[1]
\author[1,6]{Ana S. Moita}[orcid=0000-0001-9801-7617] \cormark[1]
\author[7]{Janusz Krupanek}
\author[4,8]{Yaroslav Grosu}[orcid=0000-0001-6523-1780]
\author[2]{Vitaly Zhelezny}[orcid=0000-0002-0987-1561]
\author[1]{Antonio Luis Moreira}[orcid=0000-0001-5333-5056]
\author[4,9]{Elena Palomo del Barrio}

\address[1]{IN+ Center for Innovation, Technology and Policy Research, Instituto Superior Tecnico, Universidade de Lisboa, Av. Rovisco Pais, 1049-001 Lisboa, Portugal}
\address[2]{Institute of Refrigeration, Cryotechnologies and Ecoenergetics, Odessa National Academy of Food Technologies, 1/3 Dvoryanskaya Str., Odessa 65082, Ukraine}
\address[3]{Odessa I.I. Mechnikov National University, 2 Dvoryanskaya Str., Odessa 65000, Ukraine}
\address[4]{Centre for Cooperative Research on Alternative Energies (CIC energiGUNE), Basque Research and Technology Alliance (BRTA), Alava Technology Park, Albert Einstein 48, 01510 Vitoria-Gasteiz, Spain}
\address[5]{National University "Odessa Maritime Academy", 8 Didrikhson Str., Odessa 65029, Ukraine}
\address[6]{CINAMIL - Center of Investigation, Development and Innovation of the Military Academy, Department of Exact Sciences and Engineering, Portuguese Military Academy, R. Gomes Freire, 203, 1169-203 Lisboa, Portugal}
\address[7]{Instytut Ekologii Terenów Uprzemysłowionych, Dział Badań i Rozwoju, Katowice, Poland}
\address[8]{Institute of Chemistry, University of Silesia in Katowice, Szkolna 9 street, 40-006 Katowice, Poland}
\address[9]{Ikerbasque --- Basque Foundation for Science, Mar\'ia D\'iaz Haroko 3, 48013 Bilbao, Spain}

\nonumnote{*Corresponding author: anikulin@cicenergigune.com, khliyev@ukr.net, anamoita@tecnico.ulisboa.pt}

\begin{abstract}
A new composite heat transfer fluid consisting of tetralin and fullerene has been proposed for photovoltaic thermal hybrid solar harvesting. It features a unique absorption spectrum that is capable of sharply cutting off solar energy irradiated in the range of wavelength from 300 to 650 nm, making it a perfect candidate for simultaneous harvesting of both photovoltaic and thermal components of solar energy. The proposed composite revealed outstanding stability and facile synthesize root, which are the two main obstacles for applicability of nanofluids. It was shown experimentally that the additives of fullerene to tetralin do not alter significantly it’s thermophysical properties apart from viscosity that increases moderately. Besides, tetralin/fullerene solutions show similar thermohydraulics performance to that of pure tetralin in laminar flow regime or insignificantly lower in transient and turbulent flow regimes. A new figure of merit was proposed to analyze the thermohydraulics performance that consider not only exergy losses due to the kinetic energy dissipation, but also exergy losses associated with a finite temperature difference in the heat exchanger. As a result, the proposed figure of merit indicates the decrease of the heat transfer performance of tetralin/fullerene solutions that directly proportional to fullerene concentration. The performed simulation suggests that the total energy efficiency of flat-plate photovoltaic/thermal solar collector goes up to 60.4 \% estimated according regulation (EU) No. 811/2013. Finally, life cycle analysis revealed further improvement root in view of environmental impact.

\end{abstract}

%\begin{graphicalabstract}
%\includegraphics{figs/grabs.pdf}
%\end{graphicalabstract}

%\begin{highlights}

   % \item Stable and easy-to-prepare tetralin+C$_{60}$ composite is presented
   % \item It features sharp cut off absorption spectrum desirable for PV/T splitting systems
   % \item Thermophysical properties of tetralin are insignificantly altered by addition of C$_{60}$ %\item New figure of merit (FoM) proposed for heat transfer efficiency analysis
   % \item 60.4\% of total efficiency achieved for flat-plate PV/T collector
   % \item LCA reveals best strategies for improving the sustainability of the composite

    %\item Tetralin+C$_{60}$ is easy to prepare and perfectly stable solution
    %\item Thermophysical properties of the solutions weakly change besides the viscosity that moderately increases
    %\item New figure of merit (FoM) was proposed for heat transfer efficiency analysis
    %\item 60.4 \% of total efficiency of sun power conversion was achieved for flat-plate PV/T collector according to the regulation (EU) No. 811/2013
    %\item LCA was performed for tetralin+C$_{60}$ solutions

%\end{highlights}


\maketitle

\section{Introduction}

Nowadays, many studies are devoted to the rational optimisation of different kinds of thermal equipment in order to improve their total efficiency. Among various approaches, one that attracts significant attention is related to working fluid properties tailoring. The idea proposed by \cite{choi1995enhancing}, lies in customisation of conventional working fluids properties by addition of nano-sized particles (less than 100 nm). That results in colloidal solution creation or, so called nanofluids with significantly different properties as compared to the base fluid. The field of potential applications for the nanofluids  is quite broad. For example, application of the nanofluids reveals benefits in microelectronics cooling by enhancing heat transfer coefficient \cite{al2016investigation}, that lead to temperature lowering of the thermally loaded components \cite{qi2017experimental}. Various heat sinks perform better when utilize water based nanofluids with oxides of titanium \cite{naphon2013heat}, copper \cite{peyghambarzadeh2014performance}, aluminum \cite{khoshvaght2016performance} and graphene nanoparticles.  \cite{ali2017effect}. Shell and tube \cite{karimi2020experimental},
plate \cite{kayabacsi2019experimental} and other types of heat exchangers \cite{pordanjani2019updated} have shown promising results for nanofluids application. 

An important field of nanofluids applications is solar energy harvesting as reviewed by Kasaeian et al. \cite{kasaeian2015review} and Wahab et al. \cite{wahab2019solar}. Nanofluids have been found as an efficient medium to cool down the photovoltaic cells (PVs). For example,  Karami and Rahimi \cite{karami2014heat} demonstrated reducing of the mean temperature of PVs utilizing Boehmite/water nanofluid and showed  increase of the PV electrical efficiency up to 20-37\% depending on the cooling channels configuration. Ebaid et al. \cite{ebaid2018experimental} have found that water-polyethylene glycol/TiO$_2$ and water-cetyltrimethylammonium bromide/Al$_2$O$_3$ nanofluids have superior cooling capacity of PVs as compared to the base fluids that contribute to their higher electrical efficiency.

Since the efficiency of most common and commercially available PV cells is low (around 10-20\%), a hybrid photovoltaic/thermal (PV/T) systems are evolving to enhance overall energy efficiency. PV/T systems solve at once two issues, i.e. improve the efficiency of PVs by maintaining their cells at lower temperature and utilize dissipated thermal energy that is wasted in conventional PVs. Such approach may increase total efficiency of the system up to 70\% \cite{xu2014concentration}.

Selective filtering properties of solar spectrum by some nanofluids provided an impulse for spectral splitting PV/T systems development  \cite{du2019exergy} and direct absorption solar collectors \cite{balakin2019direct} that brings another degree of freedom for solar energy harvesting systems design and optimisation.

Known heat transfer characteristics of engineered composite fluids under forced flow conditions is a relevant need in all aforementioned applications. An efficient heat transfer to the heat caring fluid is required and to a significant degree affects the energy efficiency of the system, capital costs and a payback period of equipment under design \cite{deckert2014economic}. As was aforementioned above, a substantial increase in the heat transfer coefficients (HTC) are currently reported in the literature when using nanofluids. However, the increase in nanoparticles concentration often leads to increase of the viscosity of the base fluid, that may result to several issues related to the fluid flow and to large value of pressure losses \cite{nikulin2019effect}. Also, stability of the nanofluids is an important issue, as the nanoparticles often tend to agglomerate and/or precipitate, causing large heterogeneity in the properties of the bulk fluid and precluding their commercialisation  \cite{li2009review}. Moreover, the final conclusion regarding the effects of nanoparticles additives on the base liquid heat transfer performance and overall energy efficiency is also dependent on the parameters selected to perform the analysis of the experimental results \cite{nikulin2019effect}. Thus, development and analysis of a new Fgure of Merits (FoMs) is of high priority.

For the obvious reasons, in the majority of the studies in the field of composite fluids for solar harvesting applications the most popular working fluid is water. However, nonaqueous composite working fluids may have certain benefits. For example, fullerene C$_{60}$ have a good solubility in tetralin, that further improves as the fluid temperature increases, as explained by Kozlov et al. \cite{Kozlov2007}. Since, the colloidal stability of the nanofluids is one of the most difficult obstacles to overcome and highly important for their industrial implementation, the solution of fullerene C$_{60}$ in tetralin can be considered as promising heat transfer fluid. Furthermore, the advantages of this solution are low viscosity, comparable to the viscosity of water, low saturated vapor pressure, which is an important parameter for flat-plate collectors operating at high temperatures and wide rang of working temperatures (from -35 to 207 $^{\circ}$C). 

In this study tetralin/C$_{60}$ solutions have been explored for thermal management applications in solar harvesting systems. Thermophysical properties are those that define to a large extent heat and mass transfer processes, thus a significant emphasis was given to the effect of C$_{60}$ concentration on the density, specific heat capacity, viscosity and thermal conductivity of tetralin. Heat transfer and hydrodynamic characteristics of the tetralin/C$_{60}$ solutions flowing in a cylindrical minichannel were experimentally examined in the laminar, transient and turbulent flow regimes. Light absorption of the solutions was also measured to estimate their potential for spectral splitting PV/T and direct absorption solar collectors applications. Furthermore, based on the obtained experimental data, a case study of the spectral splitting flate-plate PV/T system composed of available on the market components was simulated. Finally, life cycle analysis was performed for tetralin/C$_{60}$ solutions.

\begin{table*}  

\begin{framed}

\nomenclature[A]{$h$}{is heat transfer coefficient, [W m$^{-2}$ K$^{-1}$]}

\nomenclature[A]{$q_{l}$}{is heat flux per unit length, [W m$^{-1}$]}

\nomenclature[A]{$k$}{is liquid thermal conductivity, [W m$^{-1}$ K$^{-1}$]}

\nomenclature[A]{$c_{P}$}{is liquid specific heat capacity, [J kg$^{-1}$ K$^{-1}$]}

\nomenclature[A]{$T$}{is temperature, [K]}

\nomenclature[A]{$t$}{is temperature, [$^\circ$C]}

\nomenclature[A]{$w$}{is mass fraction of fullerene C$_{60}$, [kg  kg$^{-1}$]}

\nomenclature[A]{$q$}{is heat flux, [W m$^{-2}$]}

\nomenclature[A]{$Q$}{is volumetric flow rate, [m$^3$ s$^{-1}$]}

\nomenclature[A]{$M$}{is mass flow rate, [kg s$^{-1}]$}

\nomenclature[G]{$\mu$}{is liquid dynamic viscosity, [Pa s]}

\nomenclature[G]{$\rho$}{is liquid  density, [kg m$^{-3}$]}

\nomenclature[G]{$\nu$}{is liquid  kinematic viscosity, [m$^{2}$ s$^{-1}$]}

\nomenclature[G]{$\tau_R$}{is shear stress, [Pa]}

\nomenclature[G]{$\tau$}{is time, [s]}

\nomenclature[G]{$\dot{\gamma_R}$}{is shear rate, [s$^{-1}$]}

\nomenclature[A]{$R$}{is wire resistance, [Ohm]}

\nomenclature[A]{$Re$}{is Reynolds number $Re=\frac{v\dot D}{\nu}$, [$-$]}

\nomenclature[A]{$Nu$}{is Nusselt number $Nu=\frac{h\dot D}{k}$, [$-$]}

\nomenclature[A]{$j$}{is Colburn number $j=\frac{Nu}{Re Pr^{1/3}}$, [$-$]}

\nomenclature[A]{$Pr$}{is Prandtl number, [$-$]}

\nomenclature[A]{$v$}{is flow velocity, [m s$^{-1}$]}

\nomenclature[A]{$D$}{is internal tube diameter, [m]}

\nomenclature[A]{$L$}{is tube length, [m]}

\nomenclature[G]{$\lambda$}{is light wavelength, [nm]}

\nomenclature[A]{$a, b$}{are fitting coefficients,[$-$]}

\nomenclature[A]{$C$}{is the instrument constant related to the dimensions of the tube, [$-$]}

\nomenclature[A]{$e$}{is exergy losses, [W m$^{-2}$]}

\nomenclature[A]{$U$}{is the voltage drop, [V]}

\nomenclature[A]{$A$}{is heat transfer surface area, [m$^{2}$]}

\nomenclature[A]{$P$}{is pressure, [Pa]}

\nomenclature[A]{$f$}{is friction factor, [$-$]}

\nomenclature[A]{$R^{2}$}{is coefficient of determination, [$-$]}

\nomenclature[A]{$I_{0}(\lambda)$}{is spectral solar irradiance heat flux, [W m$^{-2}$]}

\nomenclature[A]{$I_0$}{is integral solar irradiance heat flux, [W m$^{-2}$]}

\nomenclature[G]{$\eta$}{is energy efficiency, [$-$]}

\nomenclature[G]{$\eta_{opt}$}{is glass cover transmission coefficient, [$-$]}

\nomenclature[A]{$A(\lambda)$}{is light spectral absorbance, [sm$^{-1}$]}

\nomenclature[G]{$\delta_{glass}$}{is thickness of covering glass, [m]}

\nomenclature[G]{$\delta_{insul}$}{is thickness of thermal insulation, [m]}

\nomenclature[G]{$\delta_{liq}$}{is thickness of liquid layer, [m]}

\nomenclature[G]{$\delta_{PV}$}{is thickness of PV cell used for CFD simulation, [m]}

\nomenclature[A]{$k_{glass}$}{is covering glass thermal conductivity, [W m$^{-1}$ K$^{-1}$]}

\nomenclature[A]{$k_{insul}$}{is thermal conductivity of thermal insulation, [W m$^{-1}$ K$^{-1}$]}

\nomenclature[A]{$h_{air}$}{is heat transfer coefficients to air, [W m$^{-2}$ K$^{-1}$]}

\nomenclature[A]{$h_{liq}$}{is heat transfer coefficients to liquid, [W m$^{-2}$ K$^{-1}$]}

\nomenclature[A]{$t_{out}$}{is average liquid temperature at the inlet of experimental section or collector , [$^\circ$C]}

\nomenclature[A]{$t_{out}$}{is average liquid temperature at the outlet of collector , [$^\circ$C]}

\nomenclature[A]{$t_{amb}$}{is ambient air temperature, [$^\circ$C]}

\nomenclature[A]{$\overline t_{liq}$}{is average liquid temperature along the collector length, [$^\circ$C]}

\nomenclature[A]{$\overline t_{PV}$}{is average temperature along PV cell length, [$^\circ$C]}

\nomenclature[A]{$A_{cross}$}{is collector channel cross section area, [m$^{2}$]}

\nomenclature[A]{$A_{surf}$}{is area of the collector surface, [m$^{2}$]}

\nomenclature[B]{CFD}{is computational fluid dynamics}
\nomenclature[B]{CPU}{is central processing unit}
\nomenclature[B]{FoM}{is figures of merit}
\nomenclature[B]{HTC}{is heat transfer coefficient}
\nomenclature[B]{LCA}{is life cycle analysis}
\nomenclature[B]{PEEK}{is polyether ether ketone}
\nomenclature[B]{PTFE}{is polytetrafluoroethylene}
\nomenclature[B]{PV}{is photovoltaic}
\nomenclature[B]{PV/T}{is photovoltaic/thermal}
\nomenclature[B]{THW}{is transient hot wire}
\nomenclature[B]{UV–Vis}{is ultraviolet–visible}

\printnomenclature
\end{framed}

\end{table*}

\section{Materials and methods}

This section includes the description of materials, synthesis procedure, experimental equipment and procedures used for the experimentation in this work.

\subsection{Materials, samples preparation and stability investigation}

The following materials have been used in the experiments: tetralin (1,2,3,4-Tetrahydronaphthalene, C$_{10}$H$_{12}$) (provided by Sigma Aldrich, CAS 119-64-2, purity 0.99 kg kg$^{-1}$); fullerene C$_{60}$ (provided by Suzhou Dade Carbon Nanotechnology Co., CAS 99685-96-8, purity 0.995 kg kg$^{-1}$). The solubility of fullerene C$_{60}$ in tetralin is well known and was estimated from 0.017 to 0.022 mol dm$^{-3}$ (from 0.0126 to 0.0163 kg kg$^{-1}$) at 298 K \cite{Mchedlov-Petrossyan2013}. The concentration of C$_{60}$ is important and the following cases are possible: (\textit{i}) unsaturated molecular solution of fullerene C$_{60}$ in tetralin where C$_{60}$ can be considered both as the small nanoparticles and as the large molecules \cite{Mchedlov-Petrossyan2013}; (\textit{ii}) equilibrium dissolution of C$_{60}$ in solvents where only molecular solutions are formed and cluster growth can be observed at insignificant supersaturation \cite{Avdeev2010}; (\textit{iii}) supersaturated solutions of the fullerene C$_{60}$ form clusters having tendency to precipitation. However, such clusters are unstable and can be destructed by mechanical shaking \cite{Ying1994}. Thus, all samples in this study contain the mass fractions of fullerene C$_{60}$ much below or near the saturation one.

All samples have been prepared by following steps: weighing of tetralin and C$_{60}$ using a A\&D GR-300 analytical balance with an instrument uncertainty of 0.5 mg; sonication of the tetralin and C$_{60}$ mixtures for 3 h using submersible generator UZG 13-0.1/22 (frequency 22 kHz, power 0.1 kW); 2-3 days of waiting before taking any measurement, since time to reach the equilibrium may take from several hours to days. It should be mentioned that excessive sonication does not help and may even lead to clusters formation \cite{Mchedlov-Petrossyan2013}.

\begin{figure}
\centering
\center{\includegraphics[width=0.5\linewidth]{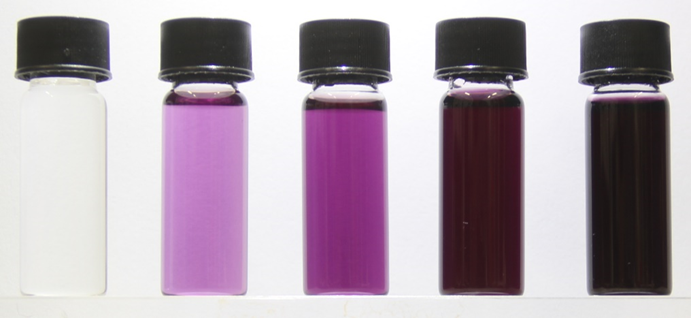}}
\caption{Samples of the tetralin and tetralin containing 0.00050, 0.00100, 0.00299 and 0.00713  kg kg$^{-1}$ of fullerene C$_{60}$ (from left to right correspondingly)} \label{Snapshot}
\end{figure}

The mass fractions of fullerene C$_{60}$ have varied from one to another experiment (see Table \ref{Concentrations}). A snapshot of the samples with various fullerene C$_{60}$ mass fractions are shown in Figure \ref{Snapshot}. 

\begin{table}
\caption{Mass fractions of fullerene C$_{60}$ in the experiments}
\label{tabular:timesandtenses}
\begin{center}
\begin{tabular}{llll}
\hline
Mass fractions, kg kg$^{-1}$ & Measurements \\ \hline
0.00501 and 0.01043 & specific heat capacity\\
0.00050, 0.00100, 0.00299, and 0.00713 & UV--vis absorption spectrum, thermal conductivity\\
0.00104, 0.00300 and 0.00655 & density, viscosity, hydrodynamic and convective heat transfer\\
\hline
\end{tabular}
\end{center}
\label{Concentrations}
\end{table}

\subsection{Measurement technique}

It is highly important to have an accurate data on the properties of the heat transfer fluid in order to estimate correctly the influence of nanomaterial additives on the single phase heat transfer and energy efficiency of the whole system. According to several studies, the thermal conductivity of nanofluids reasonably follows the expectation of theoretical models of Maxwell \cite{buongiorno2009benchmark} and Hamilton--Crosser \cite{tertsinidou2017new}. Similar trend was found for the density \cite{nikulin2019effect} and specific heat capacity \cite{o2012measurement}. However, it appears that the viscosity rarely follows the theoretical models \cite{yang2019combined}. Nonetheless, the aforementioned trends are not general rule since numerous experimental studies reported a significant deviations between experimentally measured properties of the nanofluids and predictions by the classical models. It is obvious that application of any proposed model or correlation to predict the properties of nanofluids may lead to additional errors in the final results of the heat transfer performance or energy efficiency evaluation. Thus, a fully experimental approach was followed in this study to obtain the information on the thermophysical properties of the tetralin/C$_{60}$ solutions. The measurement techniques and procedures used to characterize the density, viscosity, thermal conductivity and speciﬁc heat capacity of the nanofluids are given in paragraphs below, together with description of the procedures followed in the experiments to characterize the heat transfer in an internal flow inside a tube.

\subsubsection{Absorption spectroscopy}
One week after samples preparation the ultraviolet–visible (UV--Vis) absorbance spectrum for the pure tetraline and tetralin/C$_{60}$ solutions in the range of wavelength from 200 to 1200 nm was measured. The UV–vis spectrophotometer Cary 5000 by Agilent Technologies and standard optic cell with the optical path of 1 cm was used. The measurements have been conducted at 20 $^{\circ}$C.

In order to examine the stability of the samples, the value of absorbance at the wavelength of 630 nm for the samples containing 0.00058, 0.00177, 0.00565, 0.00687 and 0.00972 kg kg$^{-1}$ of fullerene C$_{60}$ has been measured every 2–7 days for six months. For that purpose, the UV–vis spectrophotometer Shimadzu UV-120-02 and glass optical cell with the optical path of 1.1 mm was applied.

\subsubsection{Density and viscosity}

The experimental setup for the heat transfer experiments (see Section \ref{HHT}) allows to measure simultaneously the density and viscosity of the working fluid. The density of the working ﬂuid was measured with the help of the Coriolis mass ﬂow meter, with an accuracy of $\pm$5 kg m$^{-3}$. The viscosity of the samples was measured by implementation of the capillary rheometer theory. For round cross section capillaries and tubes, the viscosity can be deﬁned from Hagen-Poiseuille equation \cite{schramm1994}.

\begin{equation}\label{mu}
\mu= C \frac{\Delta{P}}{Q},
\end{equation}
where $C=\frac{\pi D^4}{128 L}$ is the instrument constant related to the dimensions of the tube; $Q$ is the volumetric flow rate, m$^3$ s$^{-1}$; $D$ is the tube inner diameter m; $L$ is tube length, m

Shear stress $\tau_R$, Pa and shear rate $\dot{\gamma_R}$, s$^{-1}$ at the radius are defined as follow \cite{schramm1994}

\begin{equation}\label{tau}
 \tau_R = \frac{D}{4 L} \Delta{P},
 \end{equation}

 \begin{equation}\label{gamma}
 \dot{\gamma_R} = \frac{32}{\pi D^3} Q,
 \end{equation}

Usually in the capillary rheometry the volumetric flow rate $Q$ or pressure drop at the capillary $\Delta{P}$ are maintained constant. In the present study both parameters were variable and measured. The aforementioned equations (\ref{mu}-\ref{gamma}) were used only for the laminar flow regime when no heat flux was applied to the tube wall.

\subsubsection{Thermal conductivity}

\begin{figure}
\centering
\begin{minipage}[h]{0.49\linewidth}
\center{\includegraphics[width=0.3\linewidth]{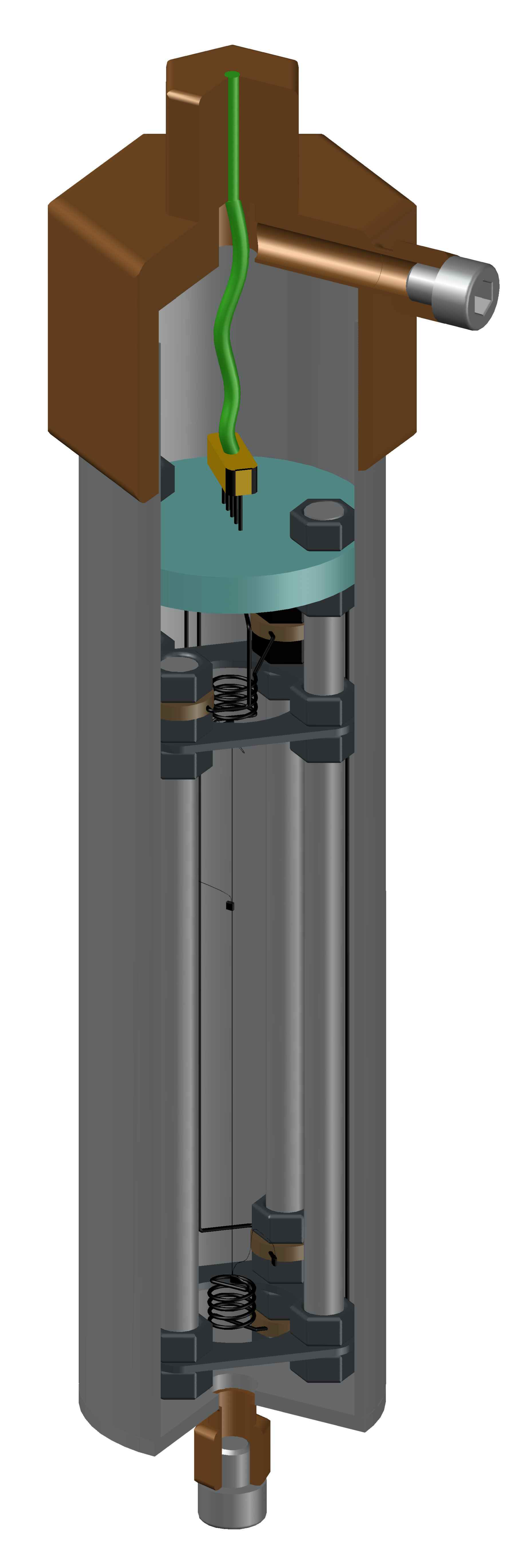} \\ (a)}
\end{minipage}
\begin{minipage}[h]{0.49\linewidth}
\center{\includegraphics[width=1\linewidth]{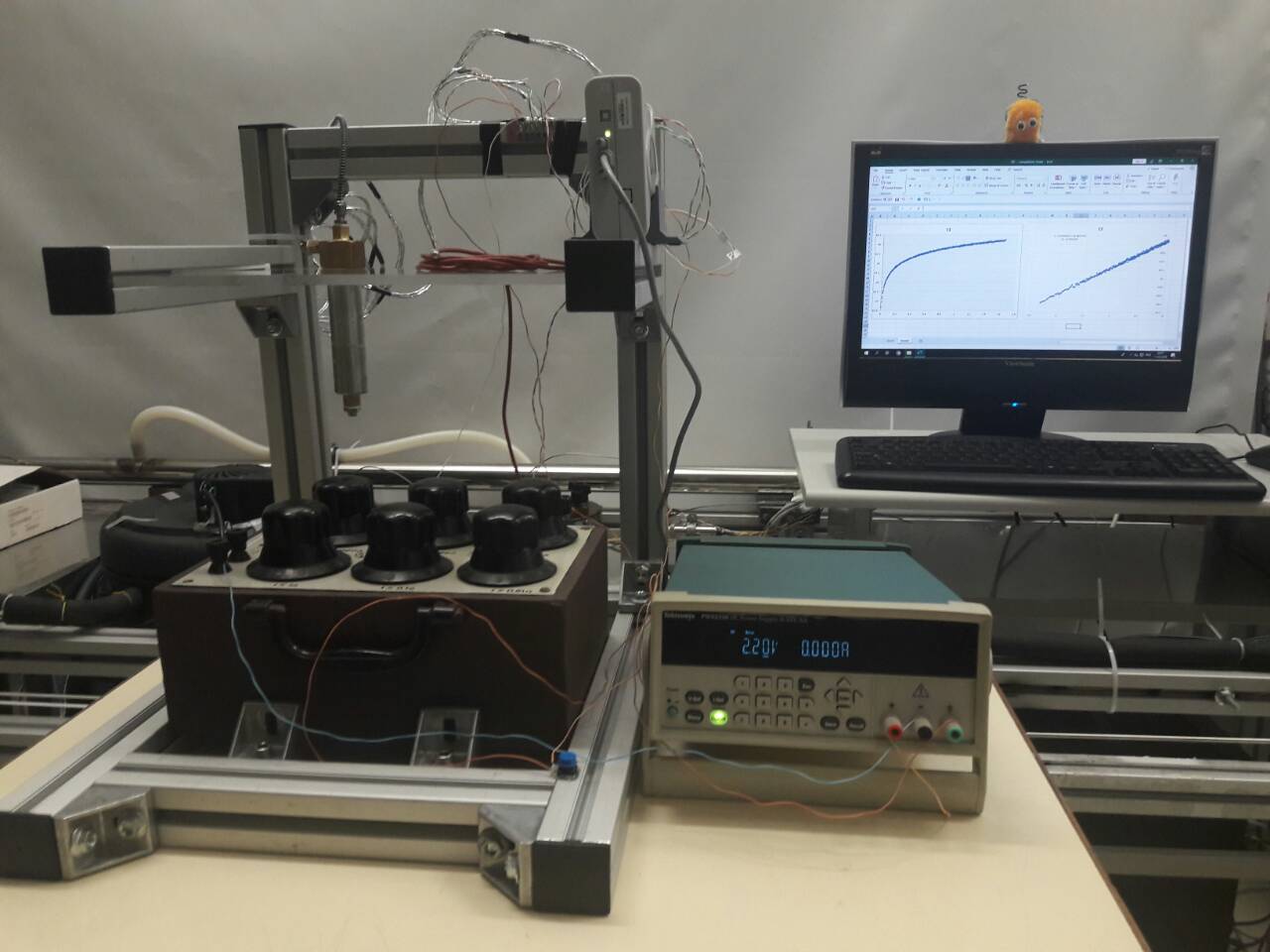} \\ (b)}
\end{minipage}
\begin{minipage}[h]{0.49\linewidth}
\center{\includegraphics[width=1\linewidth]{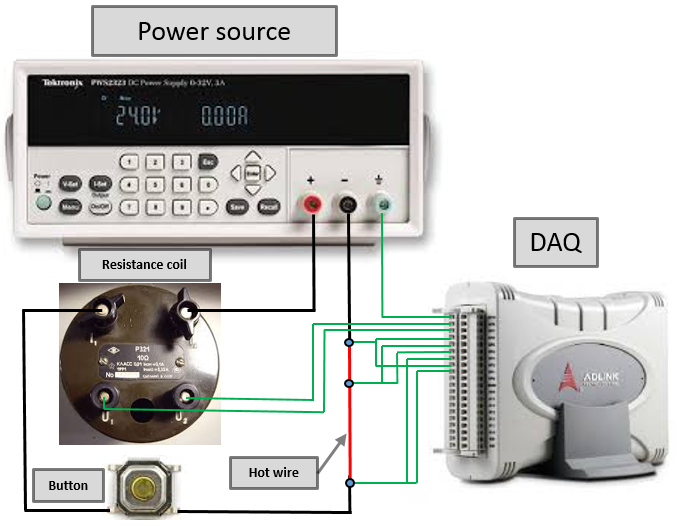} \\ (c)}
\end{minipage}
\caption{Experimental setup for the thermal conductivity measurement: (a) experimental cell drawing; (b) general view of the setup; (c) electrical scheme}\label{TC setup}
\end{figure}

To measure the thermal conductivity of tetralin and tetralin/C$_{60}$ mixtures, a custom experimental setup was created that implements the transient hot wire (THW) method. Since specification of the experimental setup was not reported before, a detailed description and validation are provided below. The experimental cell is shown in the Figure \ref{TC setup}(a). A tantalum wire with a diameter of 25 microns is used as a measuring element. The wire is divided into 2 sections, 50.5 mm and 24.6 mm long. Thermal conductivity measurements using two wires of different lengths eliminate the systematic uncertainty associated with the edge heat losses. To exclude electrical contact with the liquid, the tantalum wire was coated with tantalum pentoxide by the electrochemical method. The THW experimental cell was made following the recommendations reported in  \cite{antoniadis2016necessary}.

A thin tantalum wire (25 $\mu$m) was fixed by press welding at the ends of tantalum springs twisted out of 0.5 mm wire (4 turns with a pitch of 1 mm and a diameter of 5 mm). The ends of the springs were fixed between washers made of PEEK (polyether ether ketone), to a frame consisting of stainless steel studs with M5 thread and two perforated plates. The force of about 5 mN was applied to strain the wire.

Potential wires were also mounted at the beginning and at the end of each section of the measuring wire, made also from tantalum wire (25 $\mu$m). The ends of the power and potential wires with a diameter of 0.5 mm were led out to the upper part of the measuring cell, where they were fixed in a polytetrafluoroethylene (PTFE) washer and connected to the gold-plated collet contacts. The electrical wires were led out from the measuring cell using an electric feedthrough located in the upper part of the measuring cell. The sample is charged and drained using bolts in the upper and lower parts of the measuring cell, sealed with PTFE gaskets.
The electrical diagram of the experimental setup and its general view are shown in Figures \ref{TC setup}(b,c). A PWS2326 TEKTRONIX power supply is connected in series with the P321 reference resistance coil 10 Ohm, a short-circuit switch and a measuring wire. The temperature of the tantalum wire was measured using it´s resistance-temperature dependence and the data was acquired by the Adlink USB-2401 data acquisition system with a frequency of 322 Hz. The calculation of wire resistance was carried out according to the following formula

 \begin{equation}\label{Resist}
 R = \frac{U_{rc} U_w}{R_{rc}},
 \end{equation}
where, $U_{rc}$ is the voltage drop across the resistance coil, V; $R_{rc}$ is the resistance of the coil, Ohm and $U_{w}$ is the voltage drop on the corresponding section of the measuring wire, V.

The dependence of the wire resistance versus temperature was obtained in advance and fitted by a linear equation 

\begin{equation}\label{RvsT}
 T = aR+b,
\end{equation}
 where, $R$ is the wire resistance, Ohm; $a$ and $b$ are coefficients.

THW method is based on the solution of the Fourier´s equation for a linear heat source, which is infinitely long and thin, that dissipates heat into an infinite medium. The working equation is as follows

\begin{equation}\label{TermCond}
 k = \frac{q_l}{4\pi}\frac{d(ln(\tau))}{d\Delta T},
 \end{equation}
where, $q_l$ is the heat flux per unit length, W m$^{-1}$; $\Delta T$ is the temperature rise at corresponding time $\tau$.

\begin{figure}
\centering
\begin{minipage}[h]{0.49\linewidth}
\center{\includegraphics[width=1\linewidth]{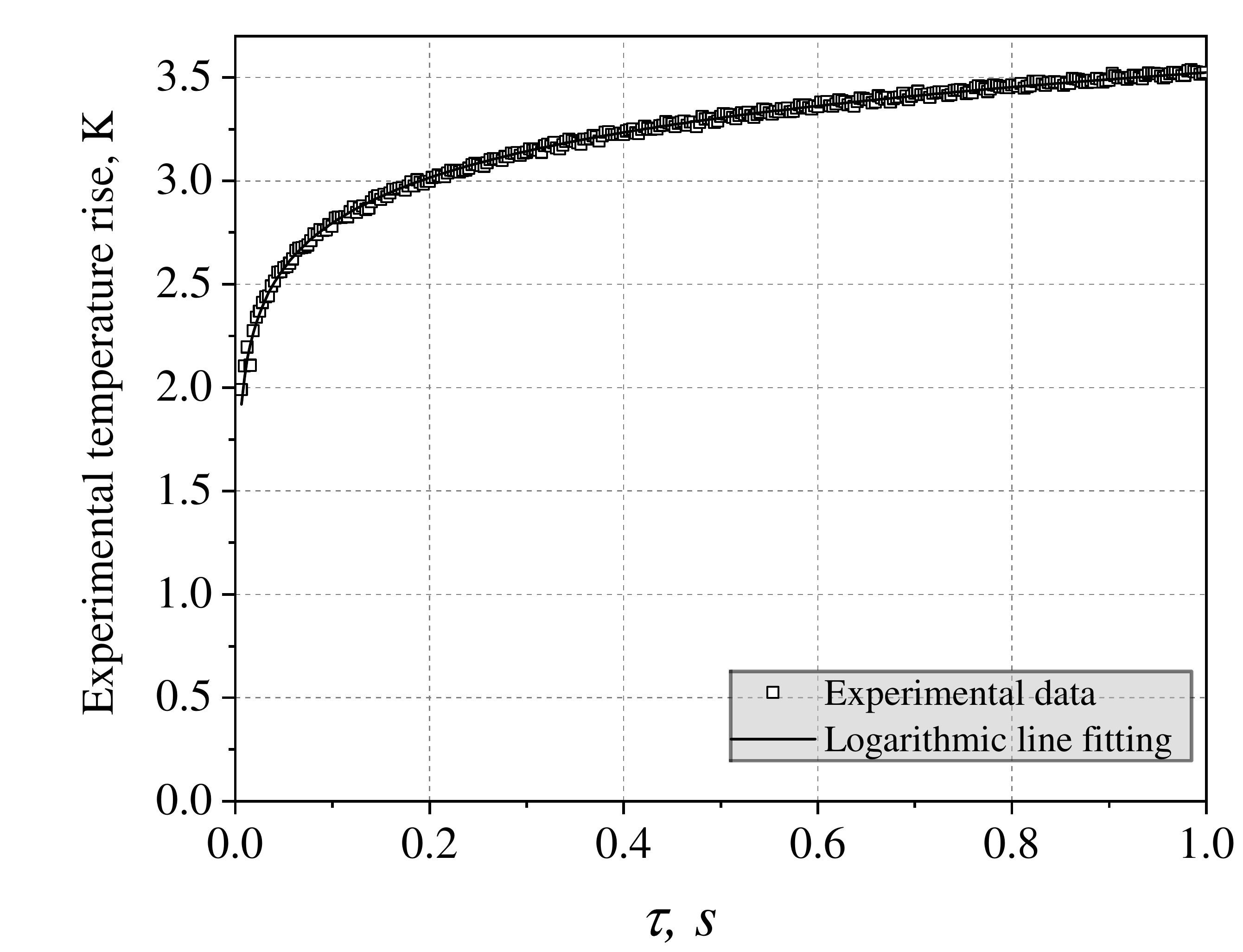} \\ (a)}
\end{minipage}
\begin{minipage}[h]{0.49\linewidth}
\center{\includegraphics[width=1\linewidth]{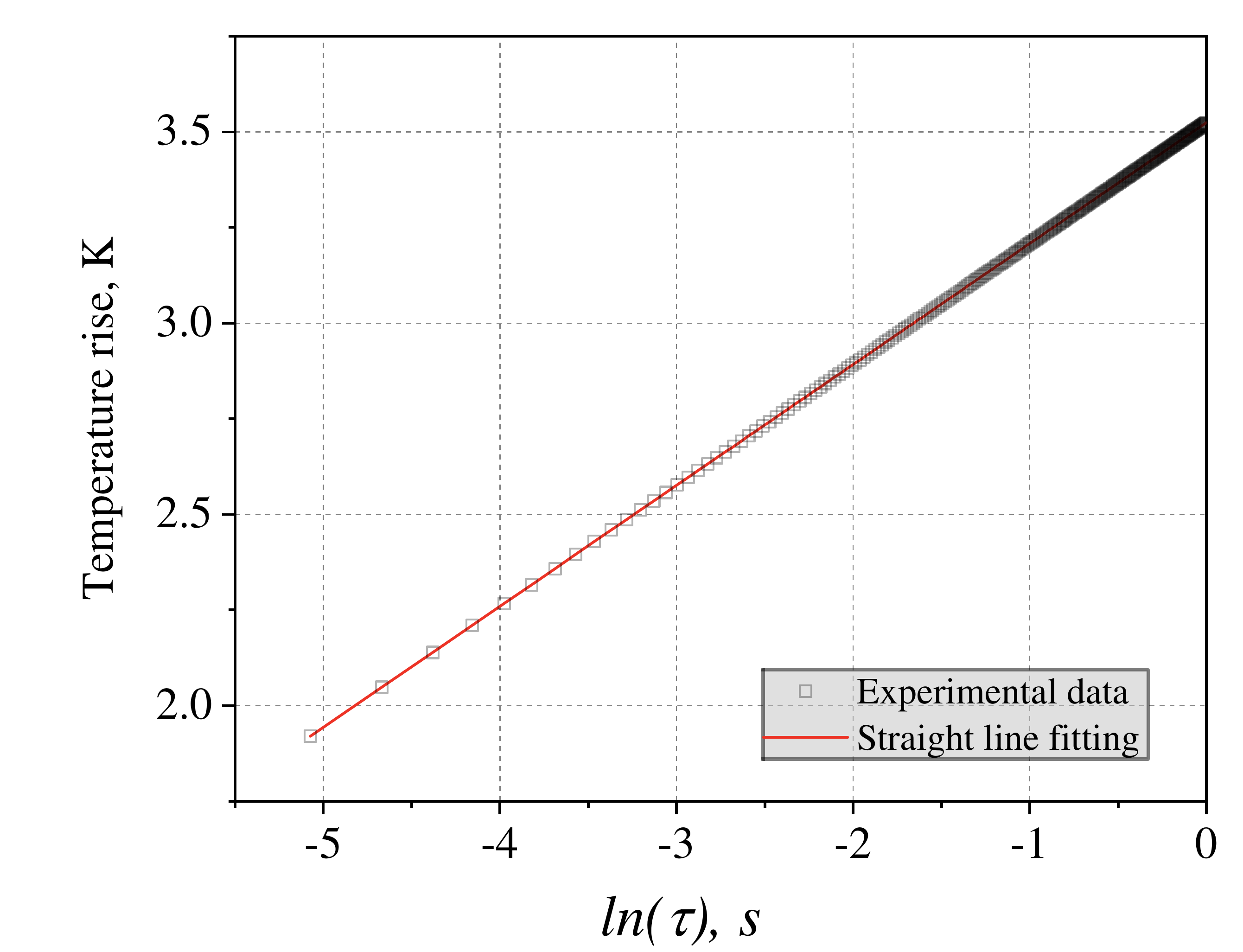} \\ (b)}
\end{minipage}
\begin{minipage}[h]{0.49\linewidth}
\center{\includegraphics[width=1\linewidth]{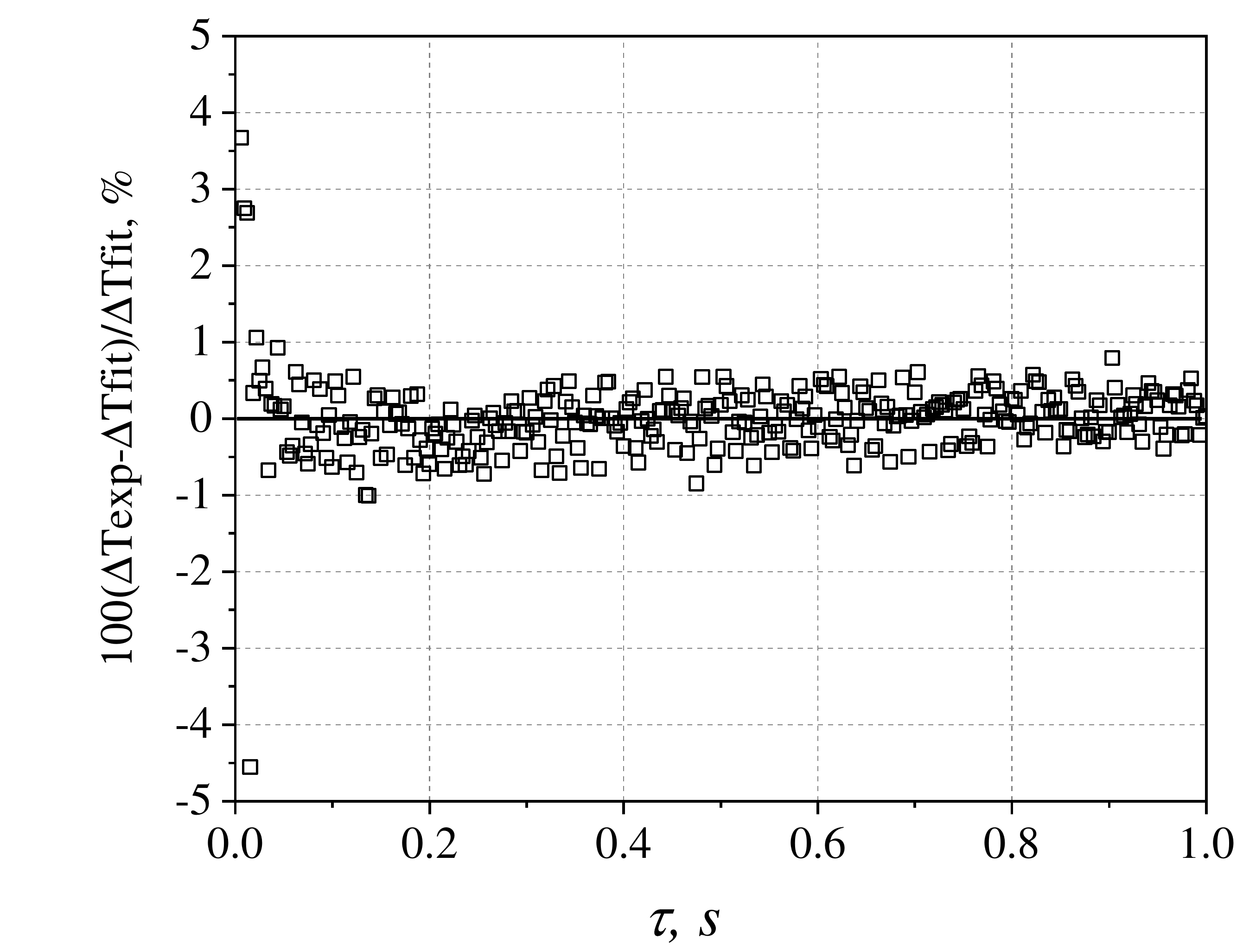} \\ (c)}
\end{minipage}
\caption{Test measurements (distilled water at t = 21.6 $^\circ$C ): (a) temperature rise vs. time; (b) temperature rise vs. logarithm of time; (c) deviations between measured values of the temperature rise and their fitting by a linear dependence}\label{TC valid}
\end{figure}

Thus, to calculate the thermal conductivity from the data obtained in the experiment, it is necessary to calculate the heat flux per unit length $q_l$ and the temperature variation $T$ over time $\tau$. The equation is valid in the region where the temperature varies linearly versus logarithm of time.
Before the experiments with tetralin/C$_{60}$ solutions, the experimental setup was validated against the prediction of a theoretical model. For this validation the thermal conductivity of distilled water and tetralin were measured. Figures \ref{TC valid}(a,b) show the data on the temperature rise of a tantalum wire immersed in water versus time and the logarithm of time, respectively, at 21.6 $^\circ$C. As can be seen from Figure \ref{TC valid}(c), the measured values of the temperature and their fitting by a linear dependence do not exhibit any systematic deviation. The deviations observed do not exceed 1$\%$ in the range from 0.1 to 1 s. This fact confirms that the experimental setup works in accordance with the mathematical description of the process.

The reproducibility of the experimental values obtained in the series of measurements for each liquid can be estimated within ± 1$\%$. The uncertainty of the experimental data on the thermal conductivity was estimated at 3$\%$, considering the random uncertainty combined with uncertainty in the temperature coefficient of the tantalum wire and other influencing factors described in \cite{alloush1982transient}.

\begin{figure}
\centering
\center{\includegraphics[width=0.5\linewidth]{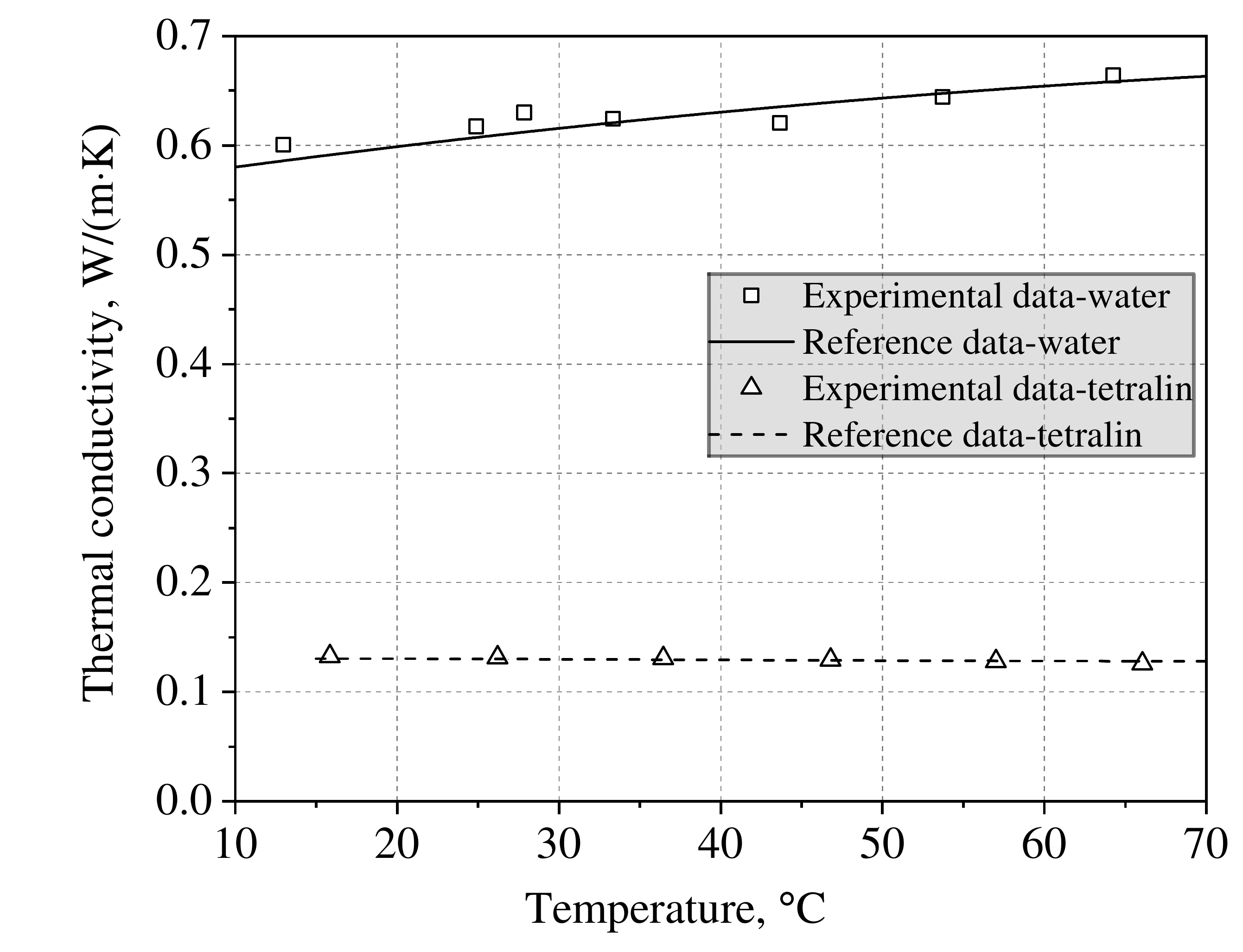}}
\caption{Comparison of the experimental and  reference data for the thermal conductivity of water and tetralin} \label{TC ExpRef}
\end{figure}

Figure \ref{TC ExpRef} compares the experimental values of the thermal conductivity for the water and tetralin against the reference data. The reference data were taken from REFPROP data base \cite{NIST2002} for water and from a handbook \cite{vargaftik1975handbook} for tetralin. Deviation between the experimental values of thermal conductivity and the reference data does not exceed the uncertainty of the experimental data for both fluids. 

\subsubsection{Heat capacity}

The measurements of the specific heat capacity for the tetralin and tetralin/C$_{60}$ solutions were performed by a monotone heating method using an adiabatic calorimeter. The scheme of the calorimeter, description of the experimental procedure and data processing are reported in \cite{zhelezny2019influence}. Main parts of the experimental setup are a container (volume 72 cm$^{3}$), copper adiabatic shields and a vacuum chamber maintained at absolute pressure of 1.3 10$^{-5}$ Pa. The continuously supplied power to the calorimeter was not higher than 0.5 W. The temperature of the container was measured by a resistance platinum thermometer. All measurements were performed with a Rigol DM3064 digital multimeter.

\subsubsection{Hydrodynamics and heat transfer}\label{HHT}

The description of the experimental setup used for hydrodynamics and heat transfer experiments and its validation was reported in detail in \cite{nikulin2019effect}. The test section is a stainless steel tube, having 4 mm in outer diameter, 0.25 mm in wall thickness and 2.4 m in length.  A 0.4 m part of the tube before the heating section serves as a settling length, to form hydrodynamically developed flow. The tube was starched using a 20 kg weight to guarantee straight and horizontal position of the test section. A high current power supply (VOLTEQ HY5050EX) was used to generate and control a constant heat ﬂux to the wall of the test section. In order to prevent heat losses to the ambient, the test section is placed in a vacuum chamber, where a dynamic vacuum of the order of 100 Pa is created by a vacuum system. The mass ﬂow rate was measured with a Coriolis mass ﬂow meter (mini CORI-FLOW M15) working in the range from 0.2 to 300 kg h$^{-1}$ with the accuracy of 0.2$\%$. Six copper-constantan thermocouples ﬁxed on the top of the tube wall are used to measure the local wall temperature along the tube. Platinum resistance thermometer installed in the ﬂow meter was used to measure the inlet temperature and a cooling system was used to maintain a desired inlet temperature in a range of 1 K. Also, a differential thermocouple and a differential pressure transducer (Omega PX2300) were used to measure the variation of temperature and pressure along the test section. All measurements were performed with a data acquisition system (RIGOL M300) in a steady state mode. The experimental procedure and data processing description can be found elsewhere \cite{nikulin2019effect}.

\section{Results and discussion}

In this section the experimental results on the optical and thermophysical properties of tetralin/C$_{60}$ solutions are reported and discussed. Results on the hydrodynamics and heat transfer of proposed working fluid were analyzed using new figure of merit. Finally, case study simulation was performed to analyze the performance of tetralin/C$_{60}$ solutions for thermal management of hybrid solar PV/T collector.

\subsection{Ultraviolet–visible absorption spectroscopy}

\begin{figure}
\centering
\center{\includegraphics[width=0.5\linewidth]{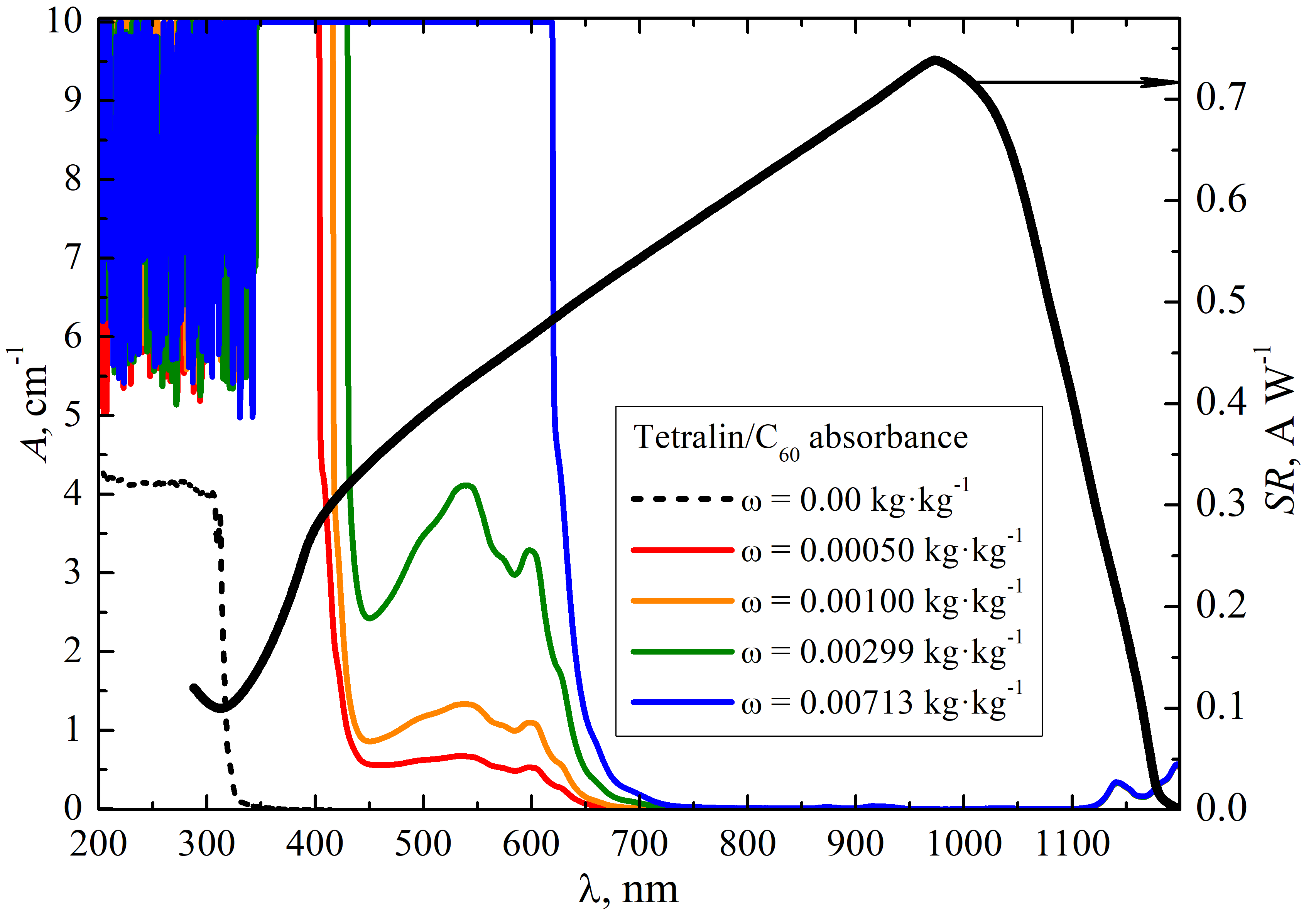}}
\caption{UV-Vis absorption spectrum of the tetralin/C$_{60}$ solutions obtained by Cary 5000 spectrophotometer  and spectral response of Sunpower silicon solar cell \cite{Looser2014} (shown by black solid line)} \label{Abs}
\end{figure}

Figure \ref{Abs} shows the absorbance spectra in the range of wavelength from 200 to 1200 nm of the pure tetraline and tetralin/C$_{60}$ solutions at the different mass fractions. It can be seen, that the absorbance of pure tetralin is close to zero in almost the entire wavelength range. At the same time tetralin/C$_{60}$ solutions show high  values of absorbance in a wavelength range approximately up to 650 nm. This property is very attractive for spectral splitting PV/T systems to harvest left part of the solar spectrum that is not used at all or used inefficiently by PV cells \cite{he2019ag}. The long-time measurements of absorbance confirmed the excellent stability to agglomeration and precipitation of C$_{60}$ in tetralin.

\subsection{Thermophysical properties}

\begin{figure}
\centering
\center{\includegraphics[width=0.5\linewidth]{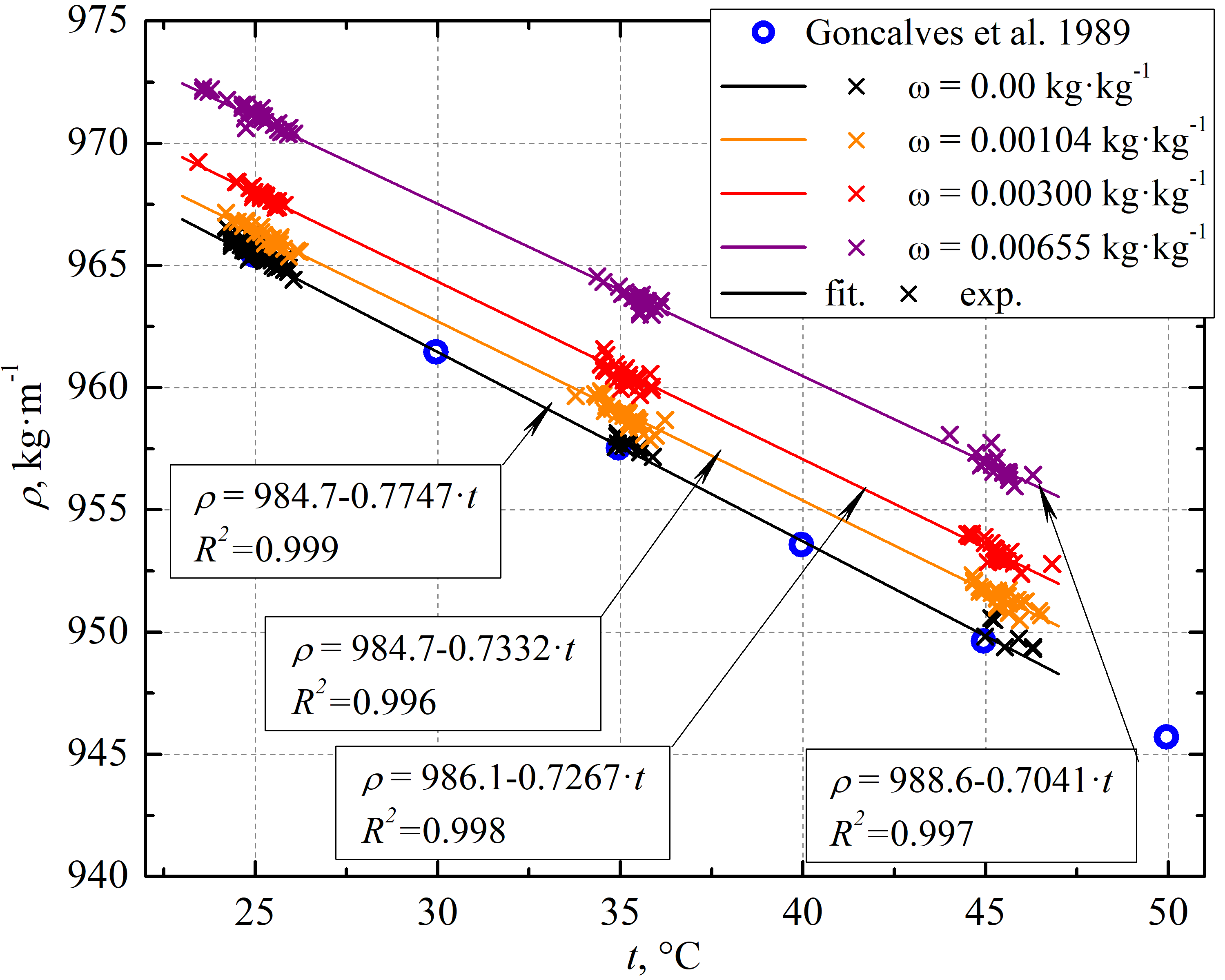}}
\caption{Temperature dependence of the tetralin/C$_{60}$ solution density at various C$_{60}$ mass fractions} \label{Density}
\end{figure}

The effect of C$_{60}$ concentration on the density of thetralin within the temperature range from 23 to 47 $^{\circ}$C, is presented in Figure \ref{Density}. The results obtained for pure tetralin clearly match with those reported by Goncalves et al. \cite{Goncalves1989}. The dependence of the density versus temperature is linear for the solutions and pure tetralin. The values of the density increase with the mass fraction of C$_{60}$. 
Obviously, the density of solutions is not significantly varies.  As can bee seen the maximum influence of C$_{60}$ on the density does not exceed 0.7\%. Fitting correlations are given in the Figure \ref{Density}.

\begin{figure}
\centering
\center{\includegraphics[width=0.5\linewidth]{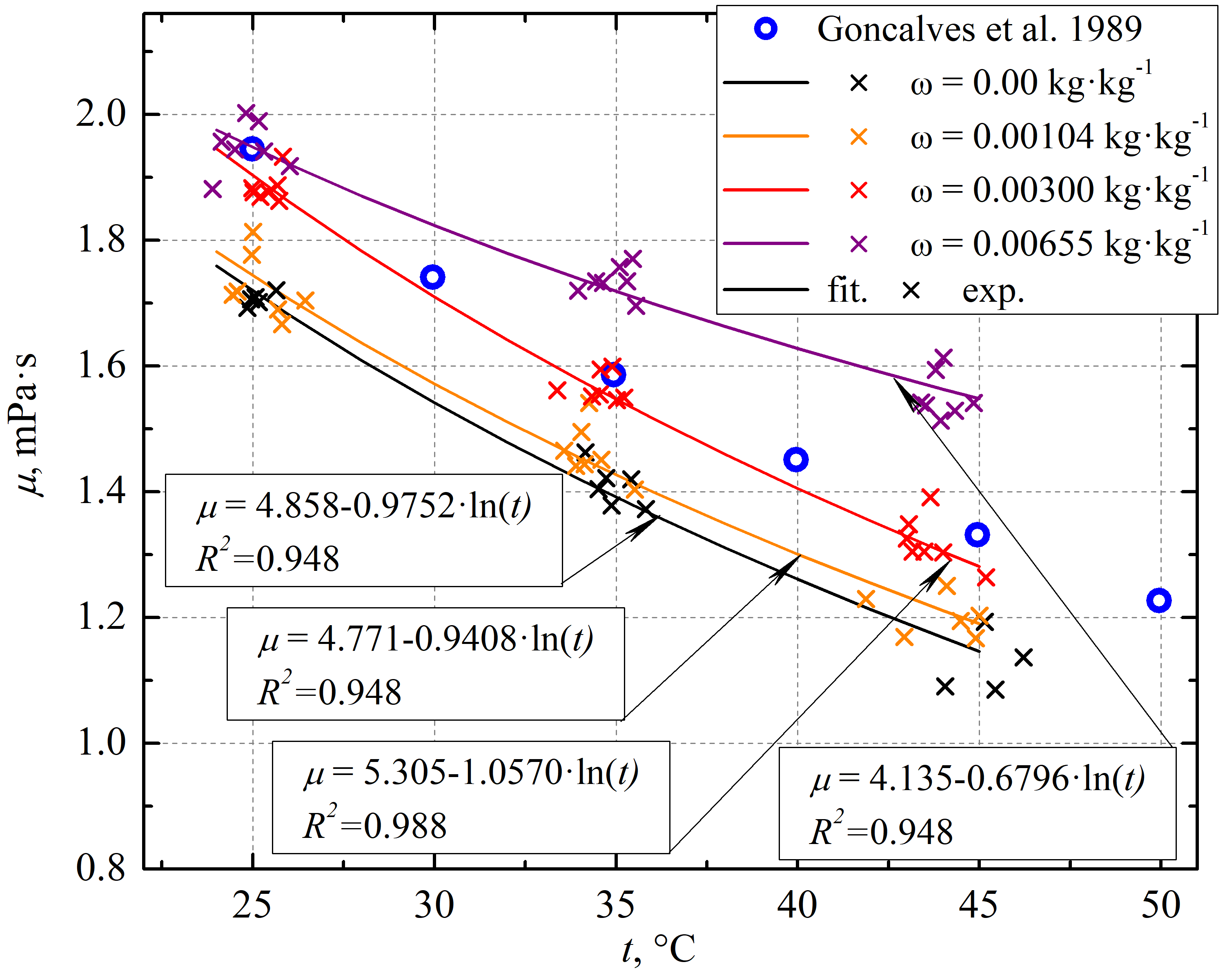}}
\caption{Temperature dependence of the tetralin/C$_{60}$ dynamic viscosity at various C$_{60}$ mass fraction} \label{Visc}
\end{figure}

The dynamic viscosity of the tetralin/C$_{60}$ solutions was measured in the temperature range from 24 to 45 $^{\circ}$C. The obtained results are shown in Figure \ref{Visc}. The viscosity data for the pure tetralin satisfactory match reference data reported by \cite{Goncalves1989}, considering the experimental uncertainty. The viscosity increases with the mass fraction of the C$_{60}$ up to 0.25 mPa s (15\%) and 0.4 mPa s (35\%) at 25 and 45 $^{\circ}$C respectively. The fitting equations depicted in the Figure \ref{Visc} have been used for the dynamic viscosity calculation, as a function of the temperature and mass fraction of C$_{60}$.

\begin{figure}
\centering
\center{\includegraphics[width=0.5\linewidth]{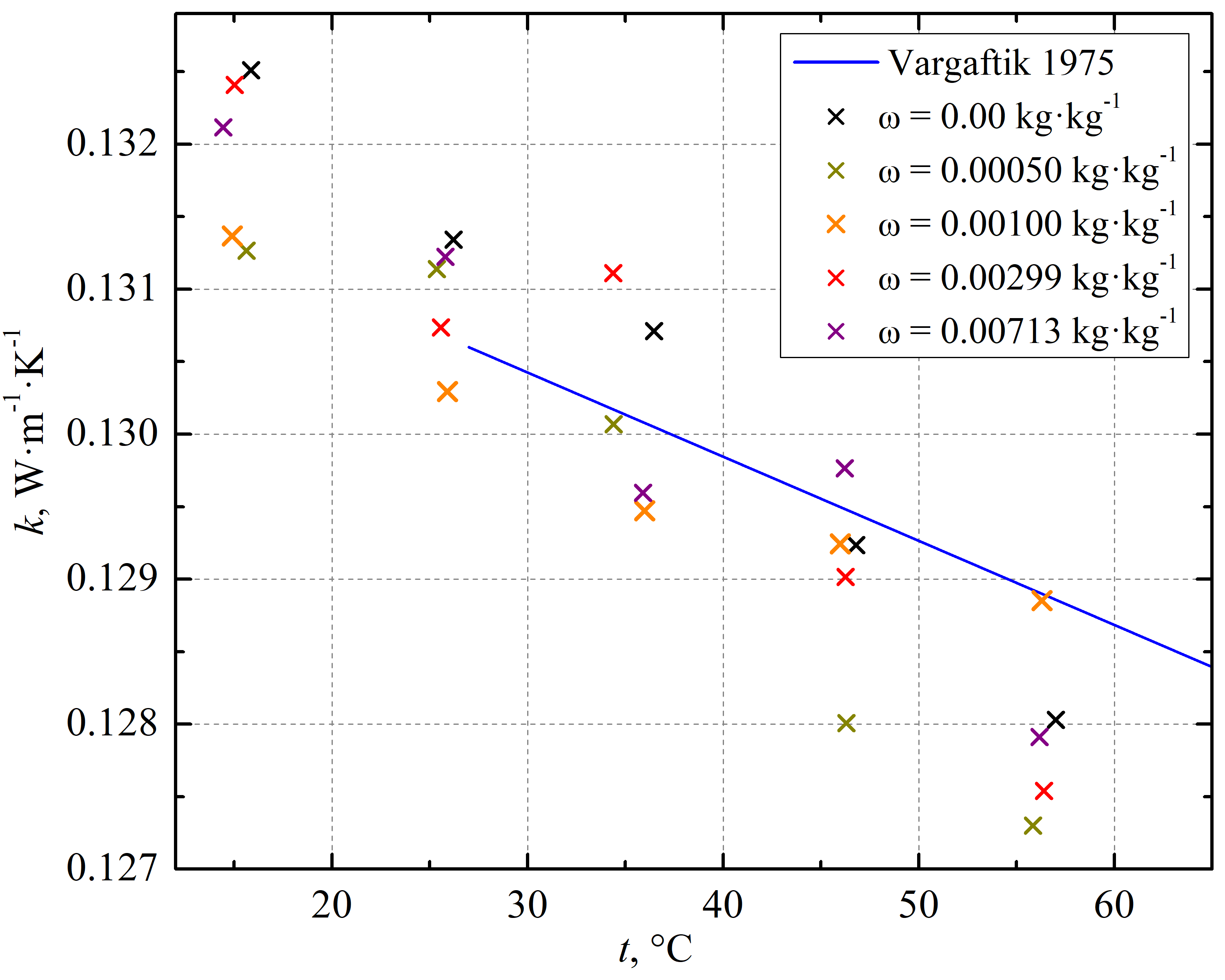}}
\caption{Temperature dependence of the tetralin/C$_{60}$ solutions thermal conductivity at various mass fractions of C$_{60}$ } \label{ThermCond}
\end{figure}

The results on the thermal conductivity that have been obtained in the temperature range from 15 to 56 $^{\circ}$C are shown in Figure \ref{ThermCond}. As can be seen, all data points spreaded within the range of 1\% and show low influence of C$_{60}$ additives on the thermal conductivity of tetralin. 

Finaly, experimental data on the specific heat capacity of the tetralin and tetralin/C$_{60}$ solutions, obtained in the temperature range from -36 to 57 $^{\circ}$C were fitted by Equations \ref{Heat capacity eq1} and \ref{Heat capacity eq2} (see Figure \ref{Heat capacity eq2}).  

\begin{equation}\label{Heat capacity eq1}
 {c_P} = A\left( {w} \right) + B\left( {w} \right) \cdot {T^3},
 \end{equation}
 
 \begin{equation}\label{Heat capacity eq2}
 A\left( {w} \right),B\left( {w} \right) = a + b \cdot {w},
 \end{equation}
 where ${c_P}$ is specific heat capacity of tetralin/C$_{60}$ solution, J kg$^{-1}$ K$^{-1}$; $T$ is temperature, K; $w$ is mass fraction of fulleren $C_{60}$ in tetralin, kg kg$^{-1}$;   $A\left( {w} \right),B\left( {w} \right)$ are concentration dependent coefficients of Equation \eqref{Heat capacity eq1}; $a$, $b$ are coefficients of Equation \eqref{Heat capacity eq2} (see Table \ref{coef of heat cap table}). 
 
 \begin{table}
\caption{Coefficients of equation \eqref{Heat capacity eq2}}
\label{tabular:timesandtenses}
\begin{center}
\begin{tabular}{lll}
\hline
Coefficients & a & b \\ \hline
$A\left( {w} \right)$ & 1284.496 & -4702.04 \\
$B\left( {w} \right)$ & 1.30334$\cdot 10^{-5}$ & 1.05198$\cdot 10^{-4}$ \\
\hline
\end{tabular}
\end{center}
\label{coef of heat cap table}
\end{table}

\begin{figure}
\centering
\begin{minipage}[h]{0.49\linewidth}
\center{\includegraphics[width=1\linewidth]{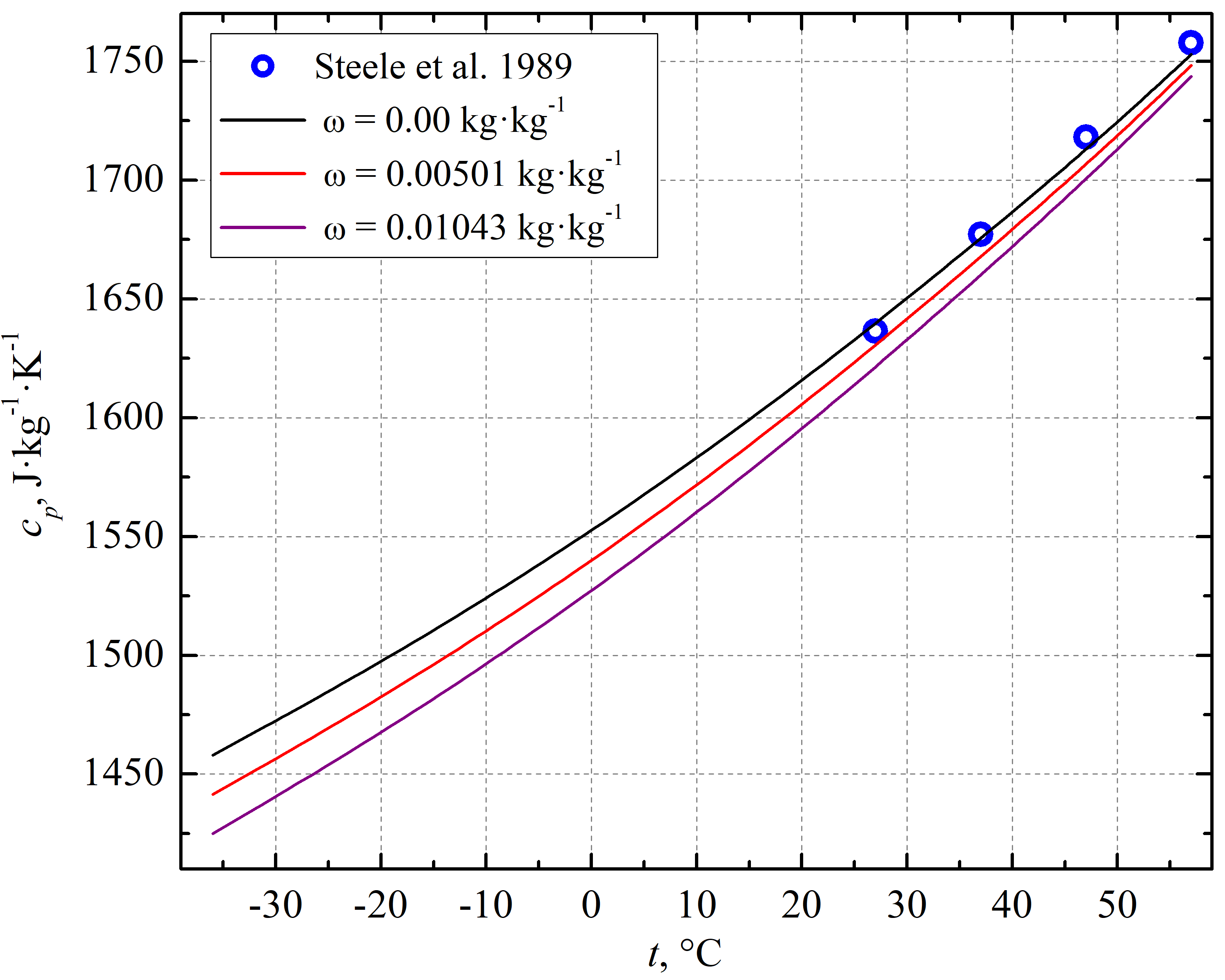}}
\end{minipage}
\caption{Temperature dependence of the specific heat capacity of  tetralin/C$_{60}$ solution at various mass fraction}\label{HC dependence liquid}
\end{figure}

The results obtained for the pure tetralin are in good agreement with reference data reported by \cite{Steele1989}. The additives of C$_{60}$ to tetralin decrease the specific heat capacity. Nonetheless, the maximum decreasing of the ${c_P}$ can be estimated in 2.1\% at the lowest temperature, while at the highest temperature the impact is less than 0.5\%. Thus, conclusion can be made, that C$_{60}$ do not significantly effect to the ${c_P}$ of the tertalin.

\subsection{Hydrodynamics}

\begin{figure}
\centering
\begin{minipage}[h]{0.49\linewidth}
\center{\includegraphics[width=1\linewidth]{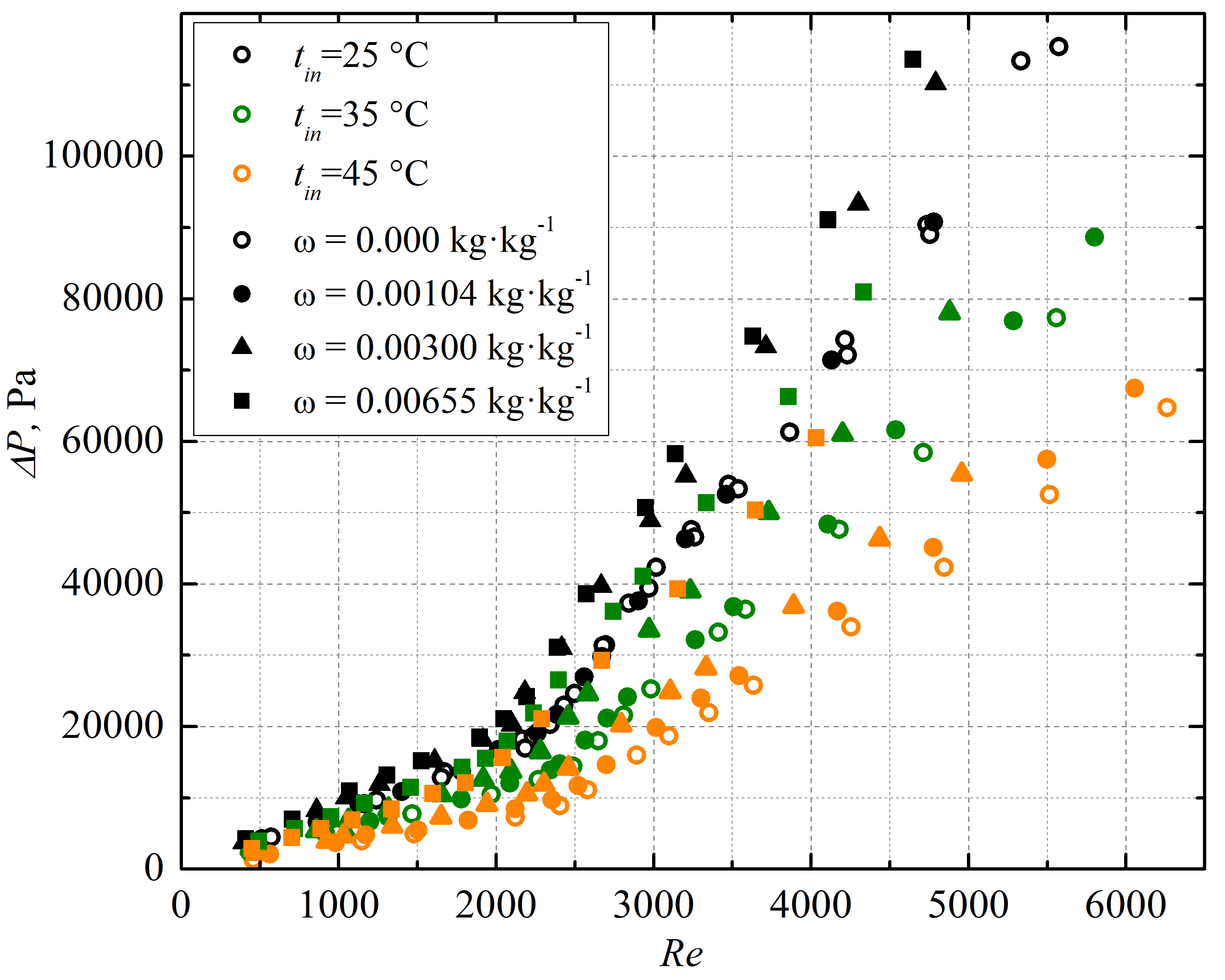} \\ (a)}
\end{minipage}
\begin{minipage}[h]{0.49\linewidth}
\center{\includegraphics[width=1\linewidth]{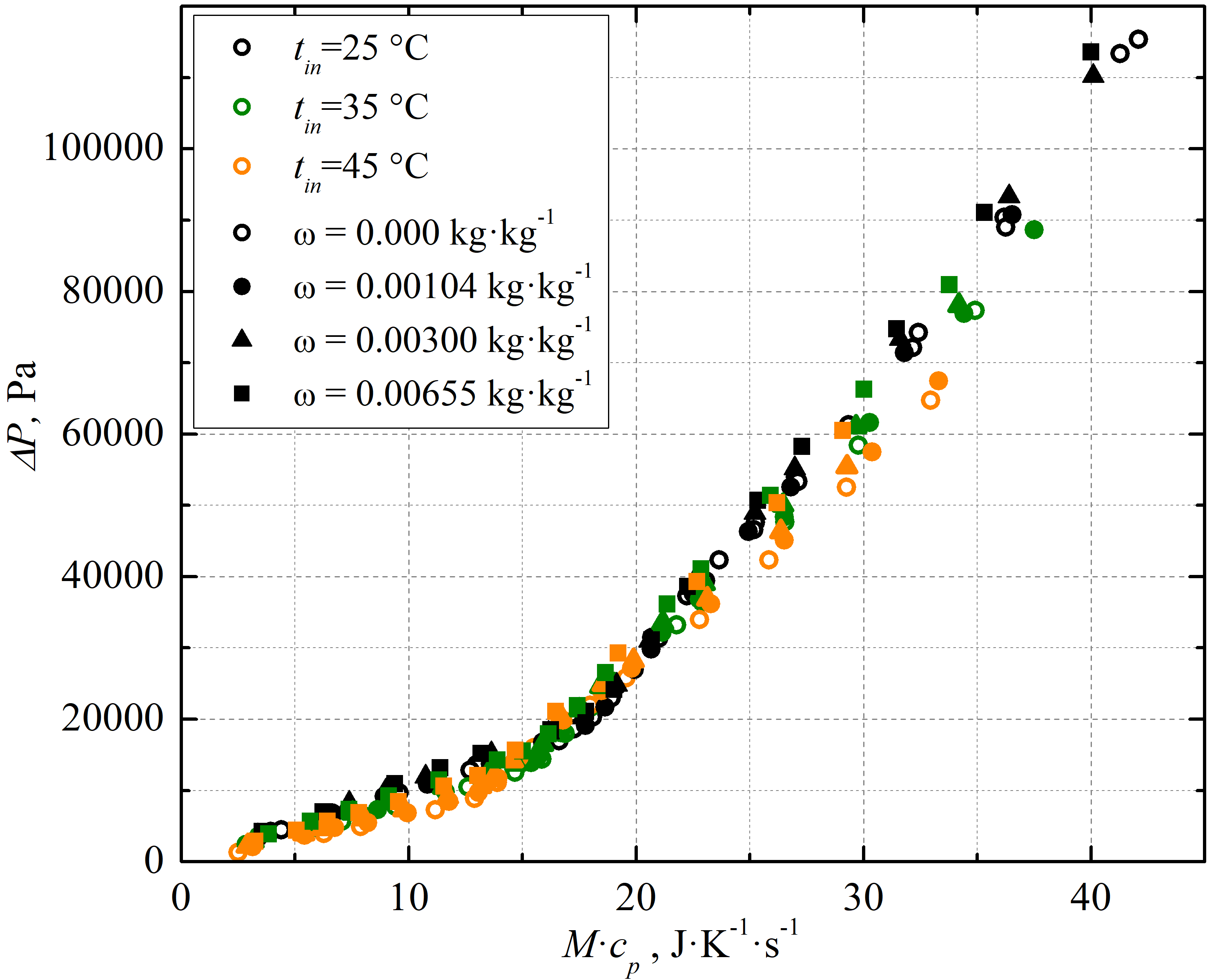} \\ (b)}
\end{minipage}
\caption{Pressure drop as function of $Re$ number (a) and $M c_{p}$ (b) at the various mass fractions and temperatures}\label{DelP}
\end{figure}

\begin{figure}
\centering
\begin{minipage}[h]{0.49\linewidth}
\center{\includegraphics[width=1\linewidth]{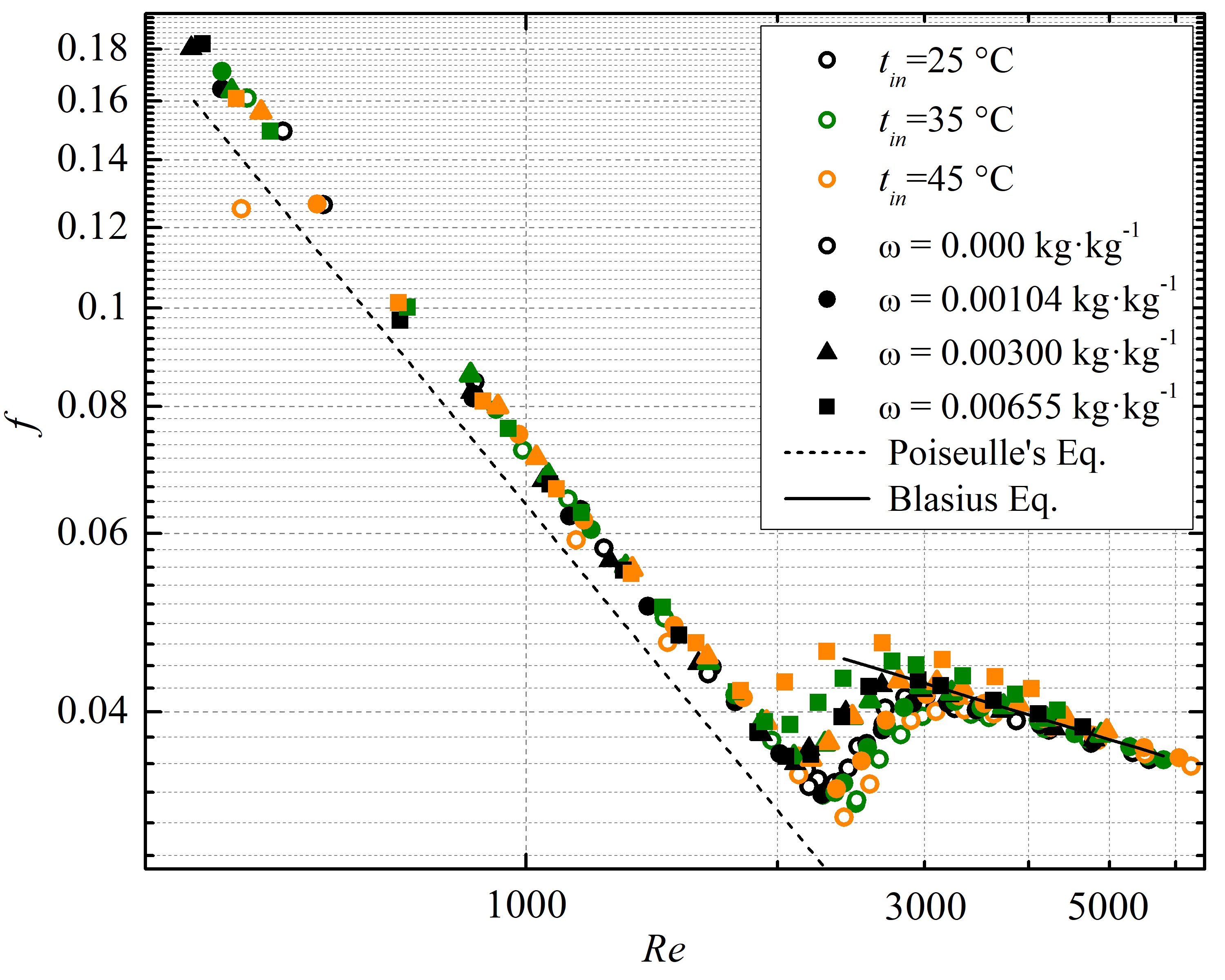} \\ (a)}
\end{minipage}
\begin{minipage}[h]{0.49\linewidth}
\center{\includegraphics[width=1\linewidth]{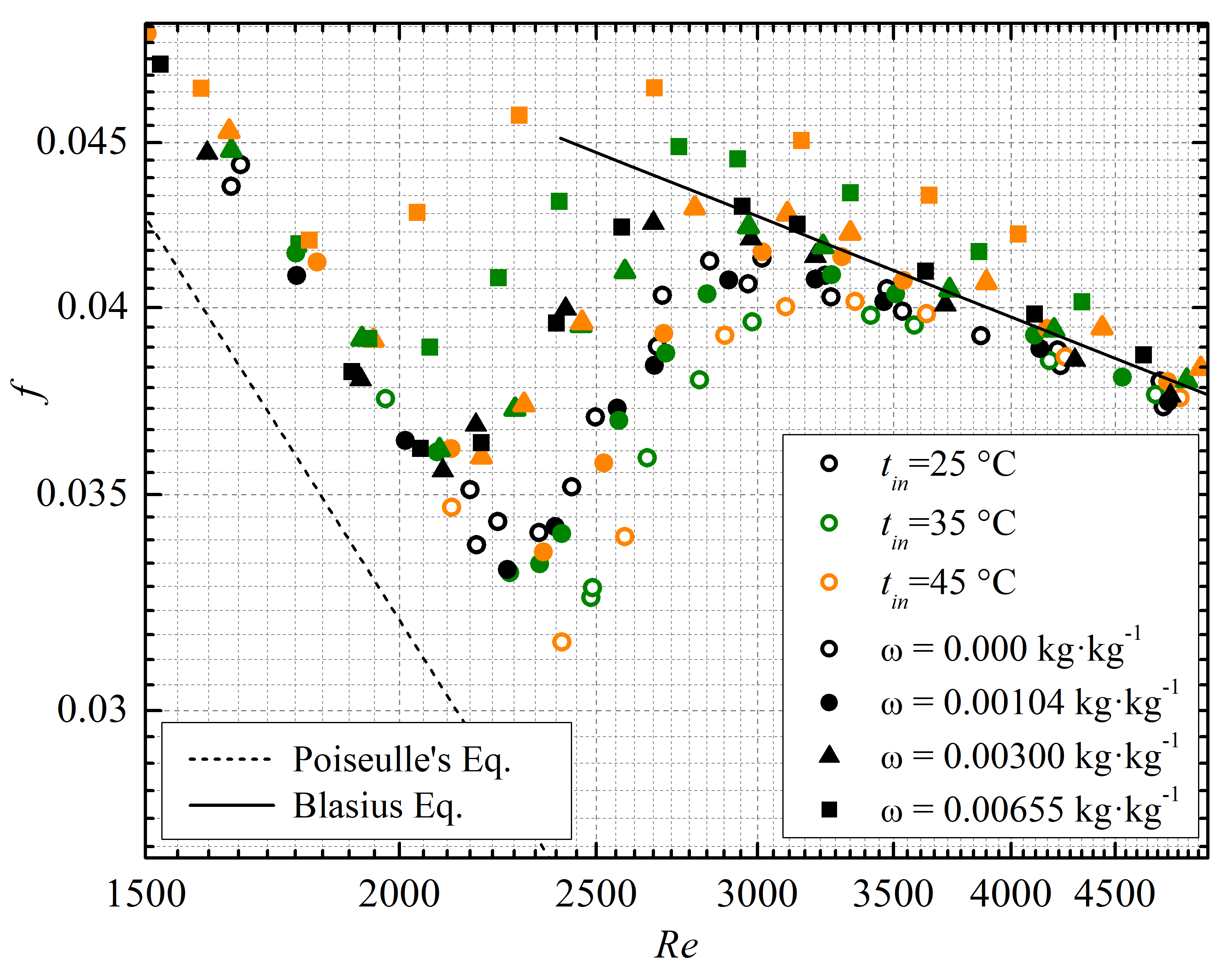} \\ (b)}
\end{minipage}
\caption{Friction factor as function of $Re$ number and mass fraction (no heat flux applied to the test section)}\label{f}
\end{figure}

The pressure drop is an important factor when analyzing an internal flows since it directly affects the required pumping power. Since the viscosity of the solutions increases with the nanoparticles mass fraction (Figure \ref{Visc}), the pressure drop is larger for the solutions with higher mass fraction of C$_{60}$ (see Figure \ref{DelP}). However, the friction factor for  all tetralin/C$_{60}$ solutions versus $Re$ number is the same as for pure tetralin (see Figure \ref{f}(a)) that agrees with the findings of \cite{nikulin2019effect}. The comparison of the experimental results with Hagen-Poiseuille and Blasius equations depicting a very good agreement (see Figure \ref{f}(a)).

The results on the friction factor (see Figure \ref{f}(b) also shown an earlier transition to the turbulent regime, i.e. the critical $Re$ number is decreasing for tetralin/C$_{60}$ solutions. This fact agrees with
the results reported in several studies for the nanofluids made of Al$_2$O$_3$ nanoparticles in  isopropanol \cite{nikulin2019effect} and water \cite{rudyak2016laminar}, as well as for multi-walled carbon nanotubes in water \cite{meyer2013influence}. Thus, the conclusion was drawn, that the solutions of C$_{60}$ in tetralin behave similar to that observed for nanofluids from the hydrodynamic viewpoint.

\subsection{Heat transfer}
\begin{figure}
\centering
\begin{minipage}[h]{0.49\linewidth}
\center{\includegraphics[width=1\linewidth]{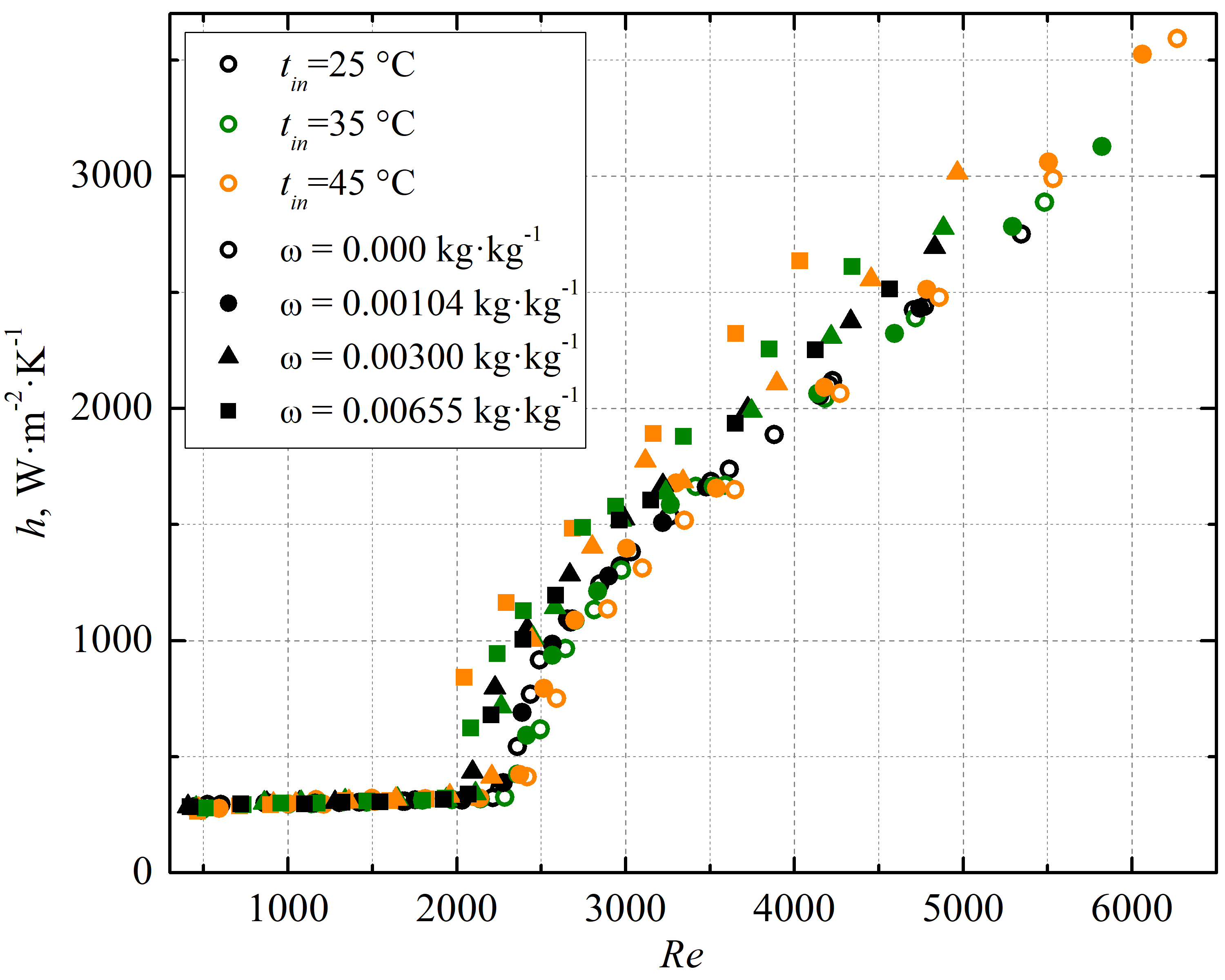} \\ (a)}
\end{minipage}
\begin{minipage}[h]{0.49\linewidth}
\center{\includegraphics[width=1\linewidth]{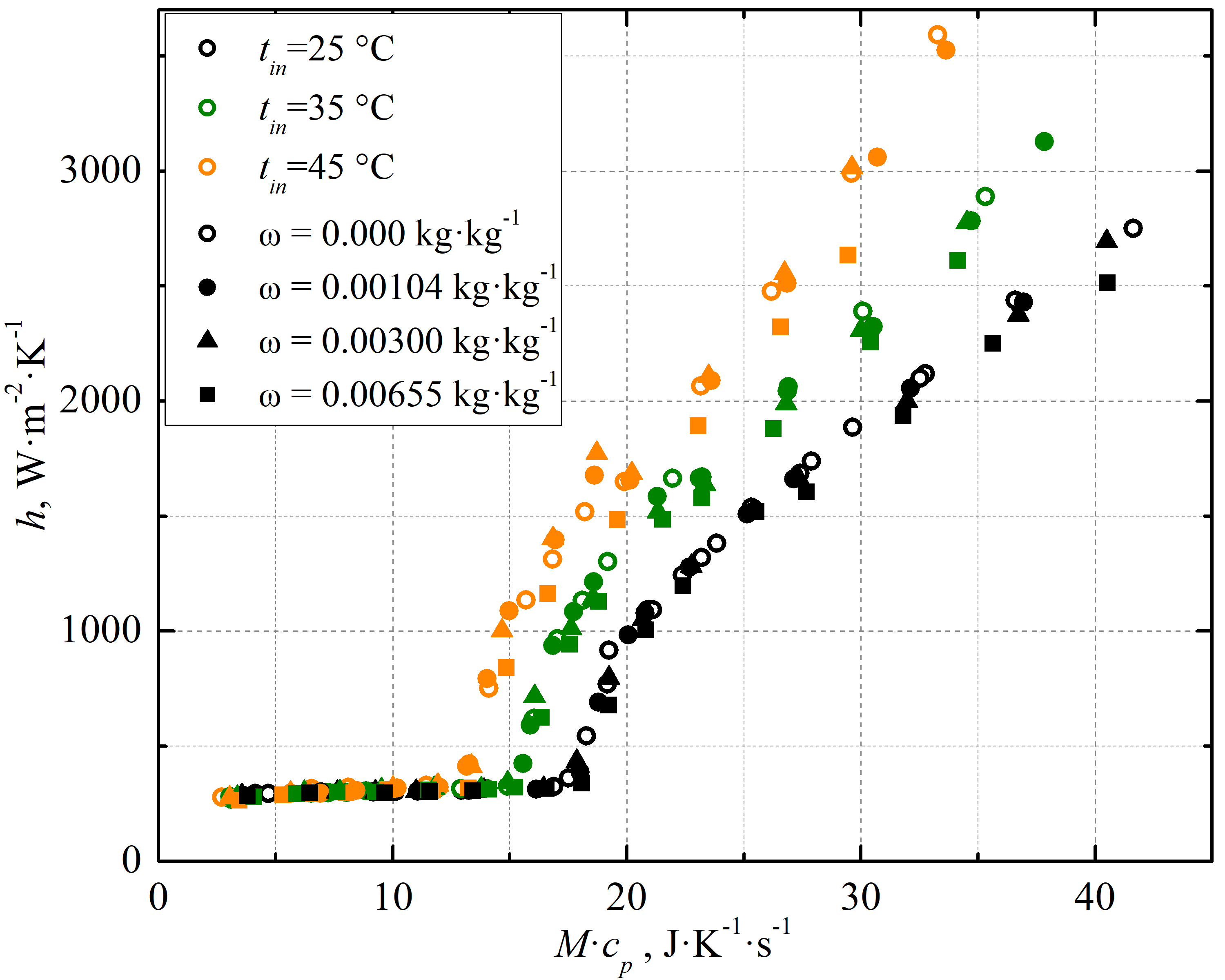} \\ (b)}
\end{minipage}
\caption{Heat transfer coefficient as function of $Re$ number (a) and $M c_{p}$ (b) at various mass fractions and inlet temperatures}\label{HTC}
\end{figure}

The experimental values obtained for the HTC as a function of the Reynolds number are depicted in Figure \ref{HTC}(a). The results reveal that in the laminar region the heat transfer coefficient is not affected by the presence of the C$_{60}$ in tetralin. However, in the turbulent region, HTC is higher for the solutions and enhances with the mass fraction of C$_{60}$ and temperature when plotted versus $Re$ number. For turbulent flow, at a $Re$ number of approximately 4000, the HTC is enhanced by 4.9, 11.2 and 30.5 $\%$ at 0.00104, 0.00300 and 0.00655 kg kg$^{-1}$ of C$_{60}$ respectively.

Since the analysed solutions are proposed to be used as the heat transfer fluids, it is more appropriate to analyse the HTC depending on the product of mass flow rate $M$ and specific heat capacity $c_{p}$, since both parameters directly determine the quantity of heat transferred in the heat exchanges. The dependence of the HTC versus $M\cdot c_{p}$ is shown in Figure \ref{HTC}(b). Here different trend can be noted, the HTC is not affected by C$_{60}$ at laminar regime but becomes slightly lower when the flow regime is turbulent.

As can be seen from the Figure \ref{HTC}(a), the additives of C$_{60}$ to tetralin lead to the earlier laminar-turbulent transition. To better identify the laminar-turbulent transition, the Colburn j-factor (Equation \ref{j-fact}) was used as a complementary parameter, as proposed by Everts and Meyer \cite{Everts2018}. 

\begin{equation}\label{j-fact}
j=\frac{Nu}{Re Pr^{1/3}},
\end{equation}

\begin{figure}
\centering
\begin{minipage}[h]{0.49\linewidth}
\center{\includegraphics[width=1\linewidth]{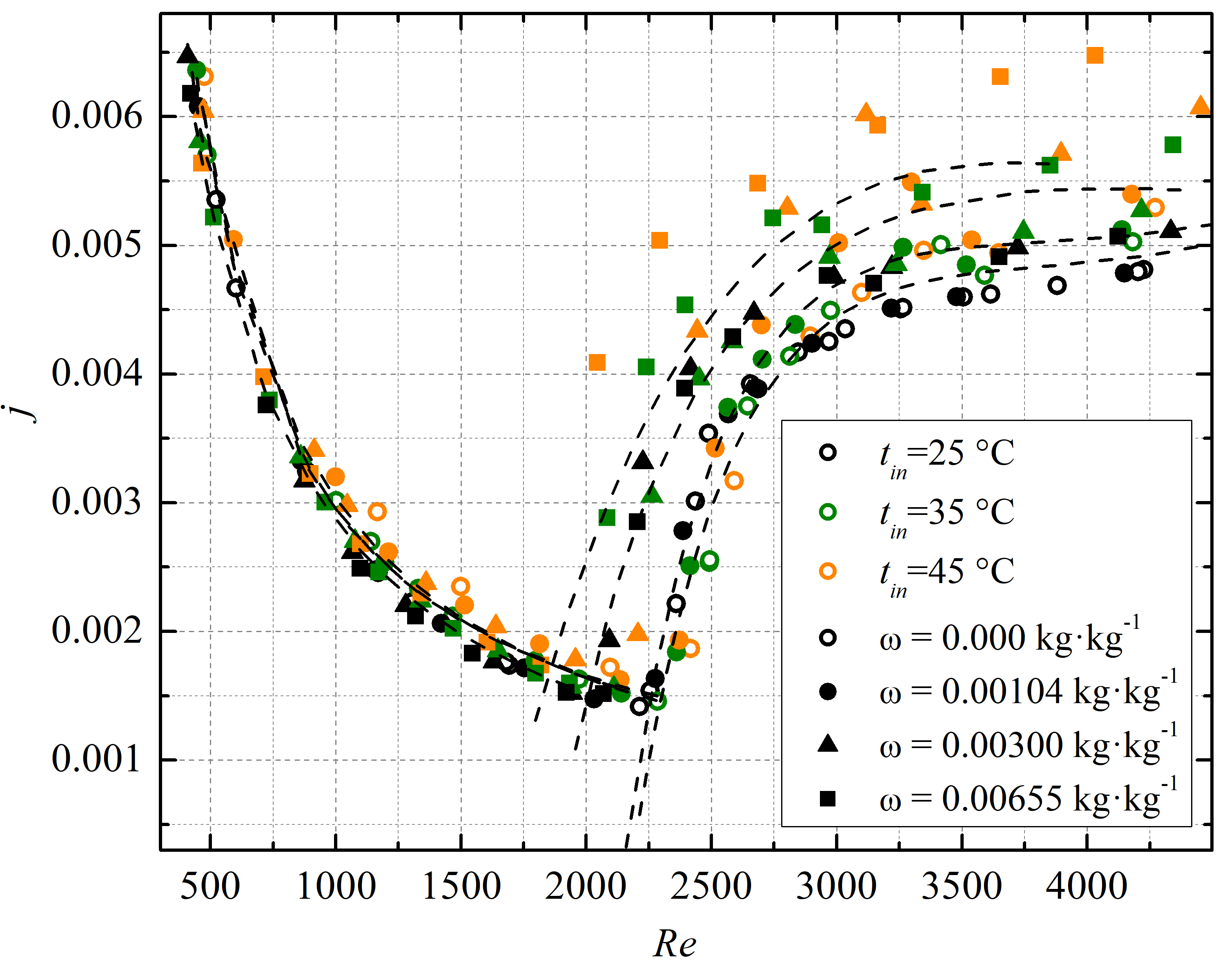} \\ (a)}
\end{minipage}
\begin{minipage}[h]{0.49\linewidth}
\center{\includegraphics[width=1\linewidth]{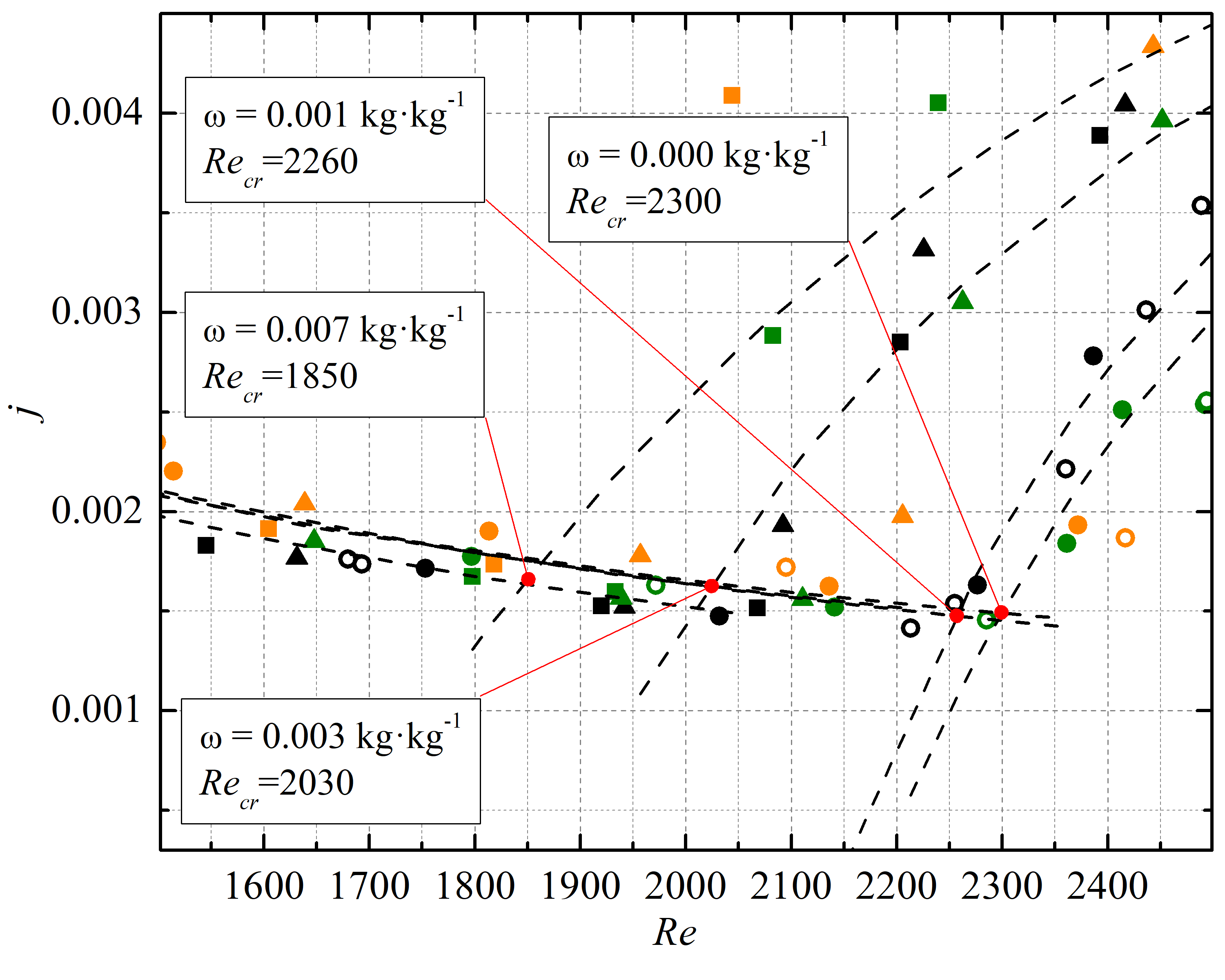} \\ (b)}
\end{minipage}
\caption{Colburn j-factor as function of $Re$ number and mass fraction (the start of the transitional flow regime are indicated by minimum of $j=f(Re)$)}\label{j-factor}
\end{figure}

The minimum of the function $j=f(Re)$ denote the ending of laminar flow existence. Looking at the Colburn j-factor as a function of the Reynolds number (Figure \ref{j-factor}), the transition from laminar to turbulent flow occurs at lower Reynolds numbers, as the C$_{60}$ mass fraction increases. The estimated critical $Re$ numbers are 2300, 2260, 2030, 1850 correspondingly for pure tetralin and the solutions having 0.00104, 0.00300 and 0.00655 kg kg$^{-1}$ of C$_{60}$. This conclusion is in line with the results obtained for pressure drop and friction factor discussed in the previous sections and may have practical importance in respect to heat transfer intensification in the heat exchangers that work in the transitional flow regime.

\subsection{Heat transfer efficiency}

Based on obtained experimental data the analysis of efficiency when utilising the tetraline/C$_{60}$ solutions as heat transfer fluid has been performed. Two criteria or figures of merit (FoM) have been considered. The simplest one is the ratio of the heat flux passing through the unit surface of the heat exchanger to the pumping power per same unit surface of the heat exchanger (see Equation \ref{FoM0}). This FoM was initially proposed to analyse the efficiency of various enhanced heat transfer surfaces \cite{Webb1981}. The use of the FoM to analyse energy efficiency of different heat transfer fluids (for example nanofluids) seems to be appropriate for nanofluids flowing in horizontal tubes \cite{Ferrouillat2011} and tested in flat plate solar collectors \cite{elcioglu2020nanofluid}. It is clear, that FoM will be correct only when the heat flux is supplied to the heat transfer fluids under comparison at equal temperature difference in the heat exchanger. Hence, the other criterion that takes into account temperature difference in the heat exchanger can be used (see Equation \ref{FoM1}). Antufiev \cite{Antufiev} has proposed similar by meaning criterion to compare thermohydraulic efficiency of heat exchangers. FoM(1) is a universal criterion that characterizes the amount of heat transferred at a temperature difference of 1 $^\circ$C and 1 W of energy spent for fluid pumping along 1 m$^2$ of the heat exchange surface. However, in authors opinion, FoM(1) is not fully suitable for practical application since it is not taking into account possible change of temperature difference because of heat transfer enhancement or deterioration when utilising different heat transfer fluids.  

Thus, a new FoM(2) is proposed here that considers two contradicting factors concerning the composite fluids \textit{(i)} exergy losses related to the power losses spent for fluid circulation and \textit{(ii)} exergy losses due to the finite temperature difference in the heat exchanger (see Equation \ref{FoM2}). Moreover, authors believe that FoM(2) may help when choosing heat transfer fluid with the aim of energy efficiency optimisation.

 \begin{equation}\label{FoM0}
 FoM= \frac {q}{e_{\Delta P}},
\end{equation}

 \begin{equation}\label{FoM1}
 FoM(1)= \frac {h\cdot{\Delta T_{1}}}{e_{\Delta P}},
\end{equation}

 \begin{equation}\label{FoM2}
 FoM(2)= \frac {h\cdot{\Delta T_{1}}}{e_{\Delta P}+e_{\Delta T}},
\end{equation}
where $q$ is heat flux, W m$^{-2}$; $e_{\Delta P}$ is exergy losses or kinetic energy dissipation of the flow related to the pressure losses, W m$^{-2}$; $h$ is HTC, W m$^{-2}$ K$^{-1}$; $\Delta T_{1}=1$ K is unit temperature difference, K;  $e_{\Delta T}$ is the exergy losses due to the finite temperature difference in the heat exchanger, W m$^{-2}$.

The following equations have been used to calculate $e_{\Delta P}$ and $e_{\Delta T}$

 \begin{equation}\label{DeltaP}
 e_{\Delta P}= \Delta P\cdot{v}\cdot(\frac{\pi\dot D^{2}}{4}),
\end{equation}
where $\Delta P$ is the pressure drop, Pa; $v$ is the flow velocity, m s$^{-1}$; $D$ is the internal tube diameter, m.

 \begin{equation}\label{DeltaT}
 e_{\Delta T}= T_{amb}\cdot{q}\cdot((\overline T_{1})^{-1}-(\overline T_{2})^{-1}),
\end{equation}

where $T_{amb}$ is the ambient temperature, K; $\overline T_{1}$ is the average temperatures of the heat transfer agent that absorb the heat, K; $\overline T_{2}$ is the average temperatures of the heat sours, K.

The average temperature of the heat transfer agent that absorb the heat was calculated as follow

 \begin{equation}\label{T1}
 \overline T_{1}=T_{in}+\frac{q \cdot A_{surf}}{2\cdot M\cdot{c_p}}
\end{equation}
where, $T_{in}$ is the inlet heat transfer fluid temperature, K; $A_{surf}$ is the heat transfer surface area, m$^{2}$.

\begin{figure}
\centering
%\begin{minipage}[h]{0.49\linewidth}
%\center{\includegraphics[width=1\linewidth]{figs/FoM(1)_new.png} \\ (a)}
%\end{minipage}
\begin{minipage}[h]{0.49\linewidth}
\center{\includegraphics[width=1\linewidth]{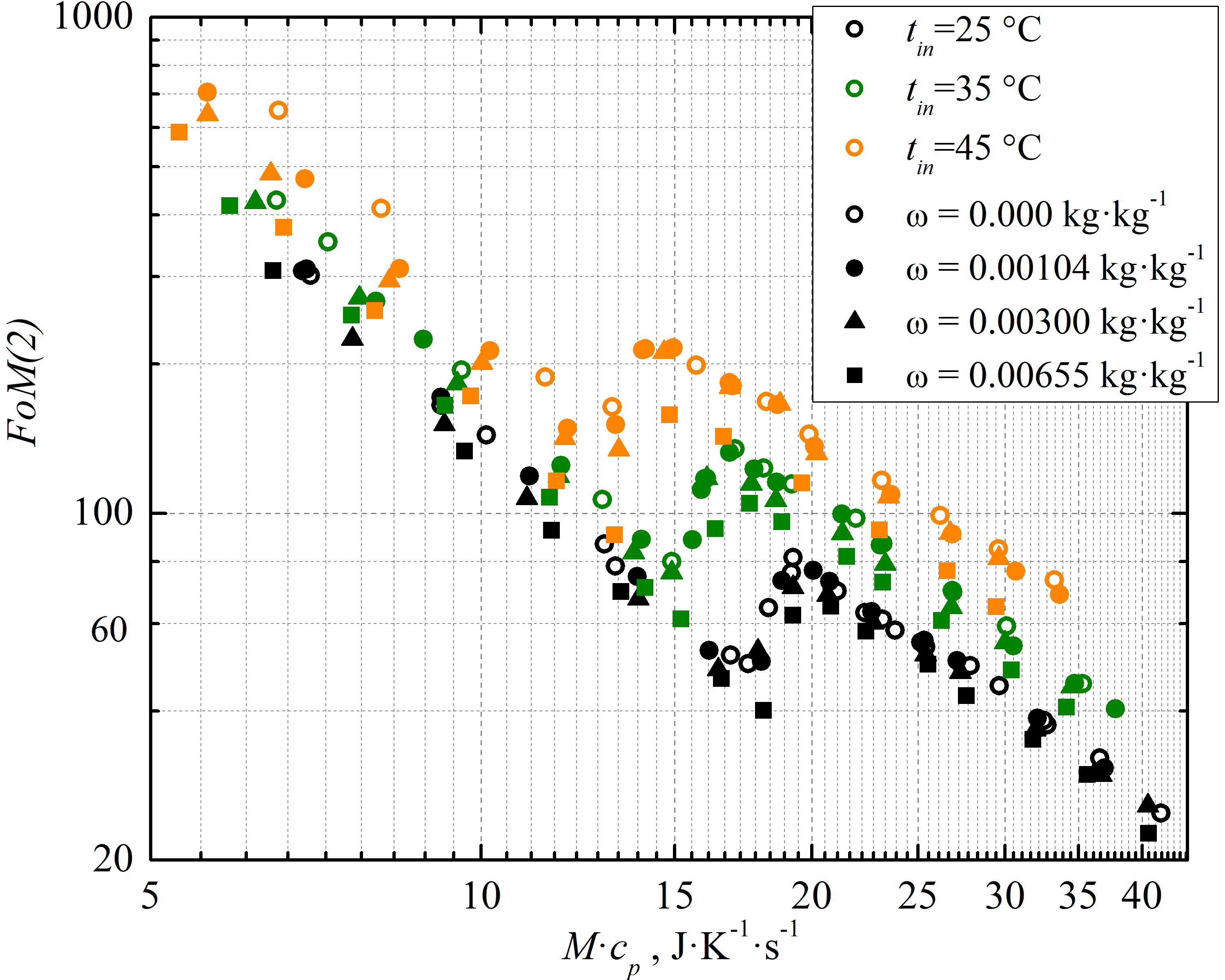} \\}
\end{minipage}
\caption{FoM(2) as function of $M c_{p}$ at various C$_{60}$ mass fractions and inlet temperatures}\label{FoM}
\end{figure}

The analysis of $FoM(2) = f (M {c_p})$ makes it possible to evaluate the qualitative characteristic of the heat transfer fluids under comparison. Obviously, the higher FoM(2) value the better performance of a specific heat transfer agent from the energy efficiency point of view. The FoM(2) was calculated at the parameters listed in the Table \ref{MoF calc} and is shown as function of $M\cdot{c_p}$ in Figure \ref{FoM}.

\begin{table}
\caption{Parameters for FoM(2) calculation}
\label{MoF calc}
\begin{center}
\begin{tabular}{ccc}
\hline
 Ambient temperature, $^{\circ}$C & Temperature of heat scour, $^{\circ}$C & Inlet temperature of heat transfer agent, $^{\circ}$C \\ \hline
 20 & 40 & 25 \\
 20 & 50 & 35 \\
 20 & 60 & 45 \\
\hline
\end{tabular}
\end{center}
\label{Param}
\end{table}

Even though, the values of FoM(2) are the highest in laminar flow, this flow regime is not practical for most of the applications. It is also notable, that the FoM(2) increases in the transitional flow regime and indicate the most energy efficient work of the heat exchanger. This is due to the sudden jump of the function $h=f(M c_{p})$ while the function $\Delta P=f(M {c_p})$ increases moderately. In the developed turbulent flow the pressure drop increases much faster than heat transfer coefficient. Thus, highly turbulent flows are less efficient from energy saving view point.

Analyzing the results shown in Figure \ref{FoM}, the conclusion was drawn, that from energy saving point of view the solutions of C$_{60}$ in tetralin have no significant difference compared to pure tetralin. Even decrease of the exergy losses due to lower temperature difference in the heat exchanger taken into account by FoM(2) do not lead to any advantages when using tetraline/C$_{60}$ solutions comparing with pure tetralin. Moreover, at the highest mass fractions of C$_{60}$ the heat transfer performance is worse compared to pure tetralin.

\subsection{Case study simulation}

\begin{figure}
\centering
\center{\includegraphics[width=1\linewidth]{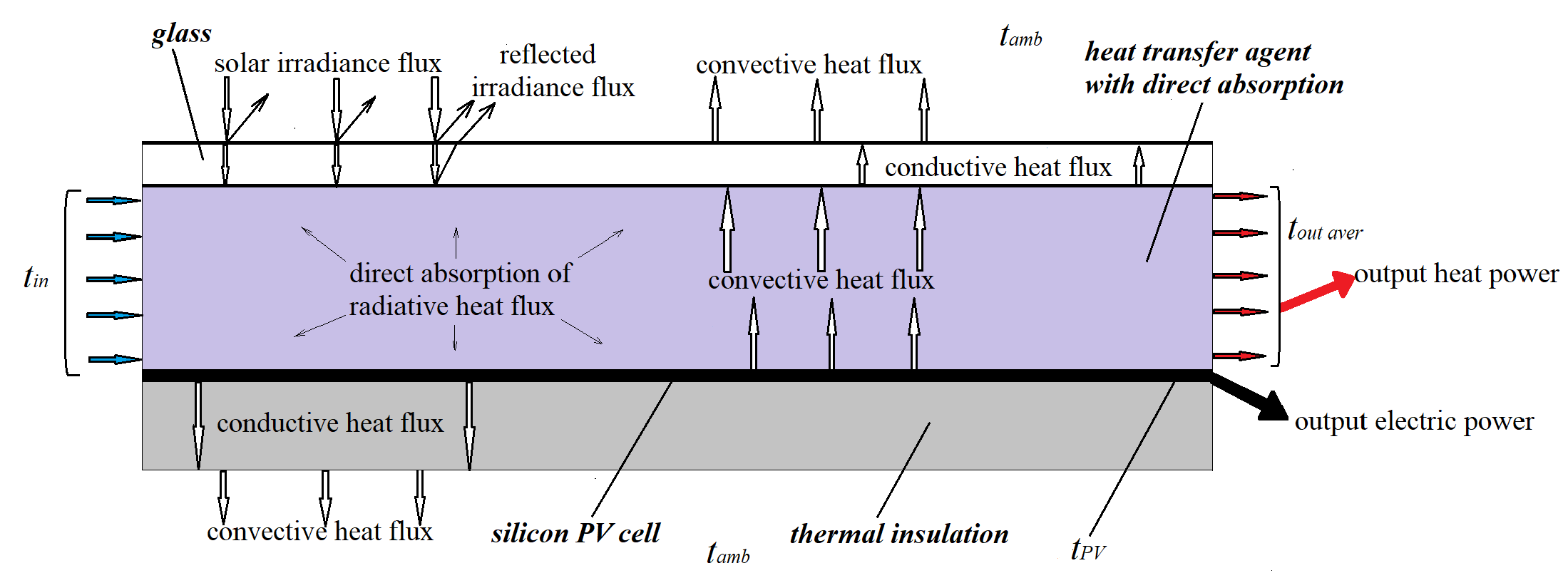}}
\caption{Energy balance diagram and modeling schematic} \label{T_PV schem}
\end{figure}

In this section the performance of tetralin/C$_{60}$ solutions as a heat transfer fluid in PV/T system was analysed. The principal scheme of direct absorption solar collector was taken from the studies of Tyagi et al. \cite{Tyagi2009} and Otanicar et al. \cite{Otanicar2010} were its high efficiency was confirmed. In the chosen for simulation PV/T system the part of the solar spectrum where the spectral response of PV cell is low is converted into thermal power. The unabsorbed part of solar spectrum by the fluid is converted by PV cell in electric power. Simultaneously, the PV cell is cooling by the fluid flowing over the PV cells. Figure \ref{T_PV schem} shows the schematic design of PV/T system that was modelled as well as the energy balance diagram.

As a prototype of combined PV/T system the flat-plate solar collectors Vitosol F series (Viessmann) was chosen where solar absorber is replaced by the silicon PV cell (Figure \ref{T_PV schem}). The input data for splitting PV/T system efficiency modelling is given in Table \ref{Input data}.

\begin{table}
\caption{Inputs data for splitting PV/T system efficiency simulation}
\label{tabular:timesandtenses}
\begin{center}
\begin{tabular}{llll}
\hline
Parameters & Value & Comments\\ \hline
Length of PV/T system & 2 m & - \\
Width of PV/T system & 0.5 m & - \\
Liquid layer thickness & 0.01 m & - \\
Channel equivalent diameter & 0.016667 m & 0.05 m channel width \\
Glass cover thickness & 0.0032 m & * \\
Ambient temperature & 25 $^{\circ}$C & - \\
Heat transfer coefficient from PV/T system to ambient & 10 W m$^{-2}$ K$^{-1}$ & - \\
Liquid velocity range & 0.01...0.05 m s$^{-1}$ & - \\
Solar transmission coefficient of glass cover & 0.96 & * \\
Thermal insulation & melamine resin foam & * \\
Thermal insulation thickness & 0.05 m & - \\
\hline
* typical value for flat-plate solar collectors Vitosol F series (Viessmann)
\end{tabular}
\end{center}
\label{Input data}
\end{table}

During the heat balance calculation experimentally obtained heat transfer coefficient in laminar flow for the solutions of C$_{60}$ in tetralin (Figure \ref{HTC}) where utilised in accordance with the similarity theory. For that reason, the channel equivalent diameter of PV/T system (Table \ref{Input data}) was used to recalculate $Re$ and $Nu$ numbers.

While modelling the convergence of the heat balance was verified by the average PV cells temperature and the temperature difference of the fluid at the inlet and at the outlet of the collector. The followings processes were considered:  

\begin{enumerate}
  \item Reflection of the sunlight from glass cover; 
  \item Volumetric absorbance of the solar irradiation by liquid (assumed constant across the fluid layer);
  \item Heat exchange between the liquid and ambient through the glass cover;
  \item Heat exchange between the liquid and PV cells;
  \item Heat exchange between the PV cells and ambient through the thermal insulation;
  \item Temperature dependence of the PV cells efficiency.
\end{enumerate}

The governing equations of the model are listed below:

Energy balance equation:
 \begin{equation}\label{EB}
 I_0 \cdot \eta_{opt}=q_{abs}+p_{TV}+q_{conv}+q_{loss1}+q_{loss2}
\end{equation}

Integral radiative heat flux transferred through liquid to the PV cell calculated by Beer–Lambert law:
 \begin{equation}\label{I_trans}
 I_{trans}=\eta_{opt} \int_{\lambda =280}^{\lambda =1200} I_{0}(\lambda) \cdot e^{- A({\lambda}) \cdot \delta_{liq}} d\lambda
\end{equation}
 
 Heat flux absorbed by the liquid:
 \begin{equation}\label{q_abs}
 q_{abs}=I_0 \cdot \eta_{opt}-I_{trans}
\end{equation}

Output electric power of PV cells:
 \begin{equation}\label{p_PV}
 p_{PV}=I_{trans} \cdot \eta_{PV}
\end{equation}

Heat flux by convection from PV cell to flowing liquid:
\begin{equation}\label{q_conv}
 q_{conv}=h_{liq}\cdot(\overline t_{PV}-\overline t_{liq})
\end{equation}

Heat flux from liquid through glass cover to air :
\begin{equation}\label{q_loss1}
 q_{loss1}=\frac{\overline t_{liq}-t_{amb}}{\frac{1}{h_{air}}+\frac{1}{h_{liq}}+\frac {\delta_{glass}}{k_{glass}}}
\end{equation}

Heat flux from PV cell through thermal insulation to air:
\begin{equation}\label{q_loss2}
 q_{loss2}=\frac{\overline t_{PV}-t_{amb}}{\frac{1}{h_{air}}+\frac {\delta_{insul}}{k_{insul}}}
\end{equation}

Heat flux consumed to heat up the liquid:
\begin{equation}\label{q_heat}
 q_{heat}=I_0 \cdot \eta_{opt}-p_{TV}-q_{loss1}+q_{loss2}=\frac{v \cdot A_{cross}\cdot \rho \cdot c_{P}\cdot ({t_{out}-t_{in}})}{A_{surf}}
\end{equation}

Thermal efficiency of PV/T system:
\begin{equation}\label{Ef_T}
 \eta_{T}=\frac{q_{heat}}{I_0}
\end{equation}

Electrical efficiency of PV/T system:
\begin{equation}\label{Ef_PV}
 \eta_{PV}=\frac{q_{heat}}{I_0}
\end{equation}

Total efficiency of PV/T system:
\begin{equation}\label{Ef_total}
 \eta_{total}=\eta_{T}+\eta_{PV}
\end{equation}

where, $I_0$ and $I_{0}(\lambda)$ are the integral and spectral solar irradiance heat flux, correspondingly, W m$^{-2}$; $\lambda$ is the light wavelength, nm; $\eta_{opt}$ is the solar transmission coefficient of glass cover; $A(\lambda)$ is the liquid light spectral absorbance, sm$^{-1}$; $\delta_{liq}$, $\delta_{glass}$ and $\delta_{insul}$ are the thickness of liquid, glass cover and thermal insulation layers, correspondingly,m; $k_{glass}$ and $k_{insul}$ are glass cover and thermal insulation thermal conductivity, correspondingly, W m$^{-1}$ K$^{-1}$; $h_{air}$ and $h_{liq}$ are heat transfer coefficients from flat surface to air and to flowing liquid, correspondingly, W m$^{-2}$ K$^{-1}$; $\overline t_{liq}=0.5\cdot(t_{in}+t_{out})$ and $\overline t_{PV}$ are the average along collector length temperatures of liquid and PV cell, correspondingly, $^\circ$C; $t_{out}$ is the average liquid temperature on collector outlet, $^\circ$C; $t_{amb}$ is ambient air temperature, $^\circ$C; $v$ is the liquid flow velocity, m s$^{-1}$;  $A_{cross}$ and $A_{surf}$ are the collector channels cross section area and area of the collector surface, m$^{2}$; $\rho$ and $c_{P}$ are the liquid density and specific heat capacity at $\overline t_{liq}$, correspondingly, kg m$^{-3}$ and J kg$^{-1}$ K$^{-1}$.

Temperature dependence of the PV cells efficiency was evaluated using the method reported by Otanicar et al. \cite{Otanicar2013} and An et al. \cite{an2017analysis}. This method requires knowledge of the incoming energy spectrum and spectral response (A W$^{-1}$)  for Sunpower silicon PV cell (with concentration ratio $C=1$) that was taken from Looser et al. \cite{Looser2014}. The incoming on the PV cell integral energy spectrum was estimated by Beer–Lambert law (Equation \ref{I_trans}) using the  standard solar spectra AM 1.5 Direct (ASTMG 173) and absorbance spectrum for solutions C$_{60}$ in tetralin (Figure \ref{Abs}). All empirical parameters for the silicon PV cell efficiency calculation were taken from An et al. \cite{an2017analysis} and listed in Table \ref{Input data}. 

\begin{figure}
\centering
\center{\includegraphics[width=0.5\linewidth]{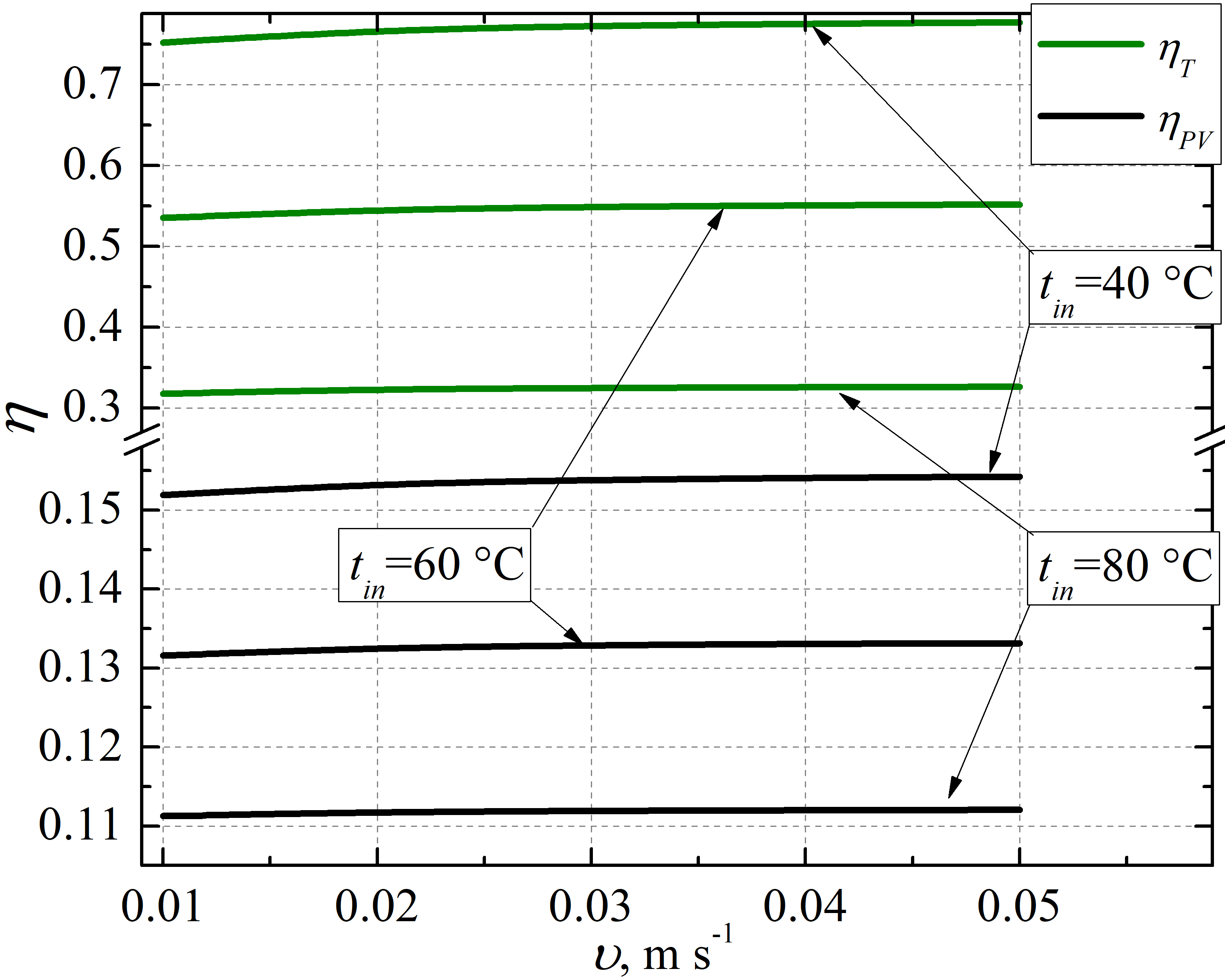}}
\caption{Dependence of the splitting PV/T system efficiency  versus liquid inlet temperature and velocity (mass fraction of C$_{60}$ in tetralin is 0.00299 kg$^{1}$ kg$^{-1}$)} \label{Effic1}
\end{figure}

\begin{figure}
\centering
\begin{minipage}[h]{0.49\linewidth}
\center{\includegraphics[width=1\linewidth]{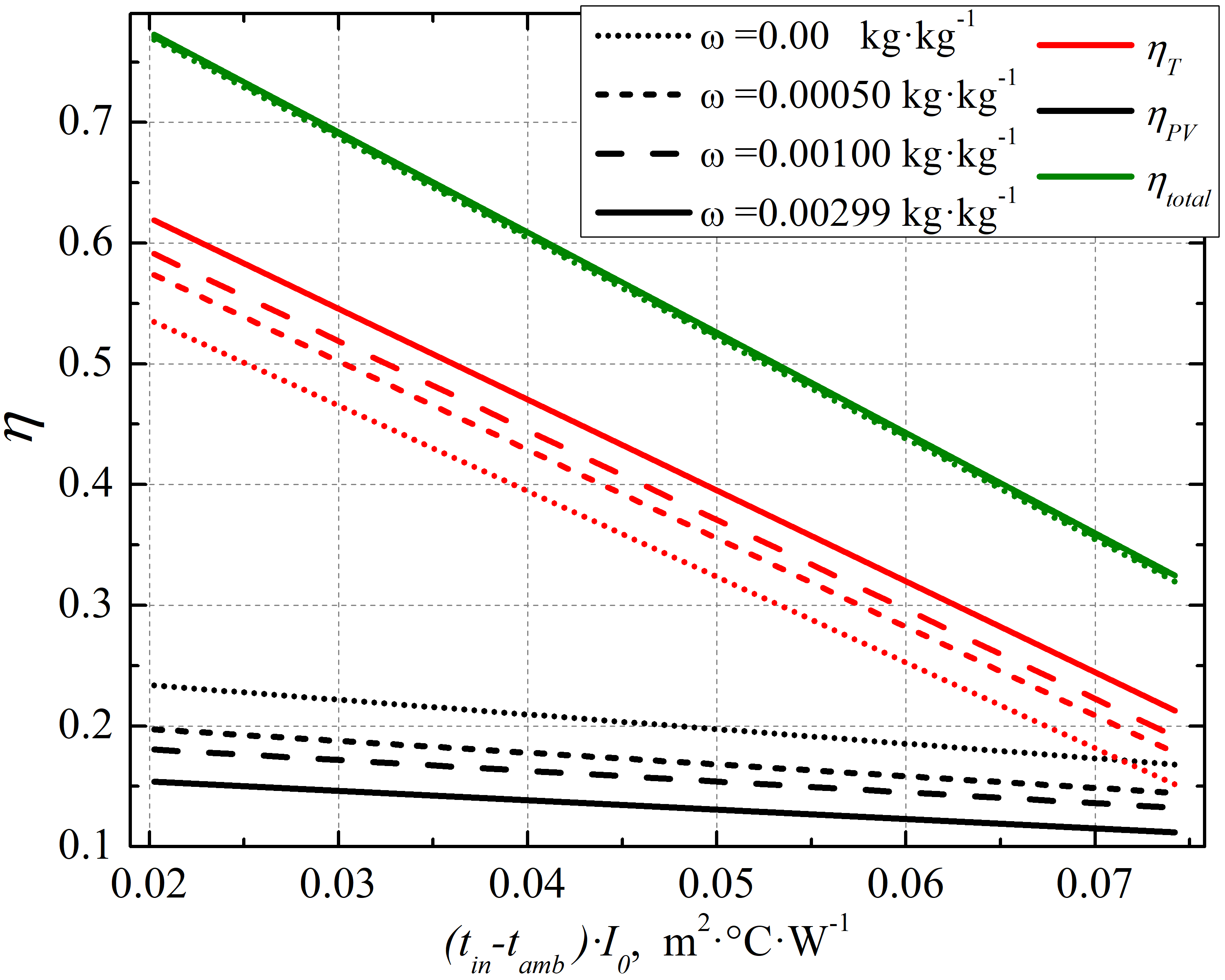} \\ (a)}
\end{minipage}
\begin{minipage}[h]{0.49\linewidth}
\center{\includegraphics[width=1\linewidth]{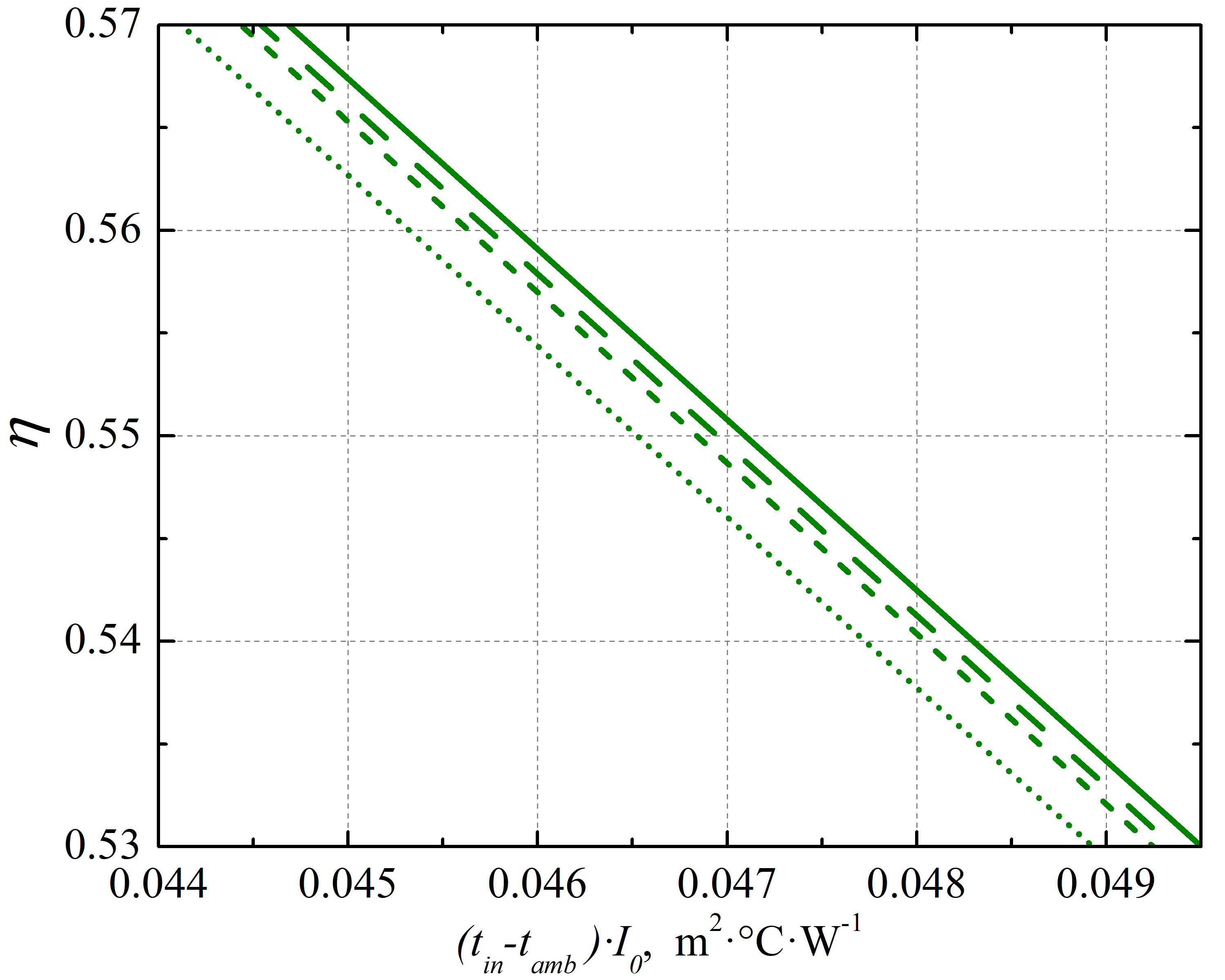} \\ (b)}
\end{minipage}
\caption{Dependence of splitting PV/T system efficiency versus $(t_{in} - t_{amb}) I_0^{-1}$ and C$_{60}$ mass fraction in tetralin (liquid velocity 0.03 m s$^{-1}$)}\label{Effic2}
\end{figure}

The results of the PV/T system efficiency modelling versus liquid flow velocities and fluid inlet temperatures at fixed ambient condition are given in Figures \ref{Effic1} and \ref{Effic2}.

As follow from Figure \ref{Effic1}(a), both thermal and PV cells efficiency slightly grow versus liquid velocity. However, above 0.03 m s$^{-1}$ this effect is insignificant. Thus, modeling of the system efficiency depending on the C$_{60}$ mass fraction was performed at the fixed fluid velocity of 0.03 m s$^{-1}$ in order to avoid unnecessary pressure losses for liquid pumping.  

The total efficiency of the PV/T system slightly grow (up to 0.5 \%) over the mass fraction of C$_{60}$  (see Figure \ref{Effic2}(b)). At the same time, the efficiency of PV cell decreases substantially for the solutions with higher content of C$_{60}$. The opposite trend is observed for thermal efficiency, that increases versus particles mass fraction. Therefore, solar energy transformation is controlled by the C$_{60}$ concentration proportionally between electrical and thermal energy. However, considering the value of electricity compared to thermal energy, high cost of fullerene C$_{60}$ for today, increase of the viscosity and pressure loses as well as lower heat transfer efficiency of the solutions with higher content of C$_{60}$ additives, it is expedient to use the solutions with lower fullerene load or pure tetralin.

The efficiency of the PV cell decreases slowly in the examined temperature range (see Figure \ref{Effic2}(a)). Thermal efficiency of the system sharply decreases versus inlet temperature on the liquid, that is associated with the heat looses through glass cover and is typical for all flat-plate solar collectors.

According the Regulation (EU) No 811/2013, the solar collector efficiency means the efficiency at a temperature difference between the solar collector and  surrounding air of 40 K at the total solar irradiance of 1 kW m$^{-2}$. For comparison, the value of $\eta_{total}$ for flat-plate solar collector Vitosol F series (Viessmann) at mentioned parameters lie in the range 57...63.9 \% (depend on the absorber coatings). At the same time, the value of $\eta_{total}$ calculated by Eqs. \ref{EB} to \ref{Ef_total} for pure tetralin velocity 0.03 m s$^{-1}$ is equal to 60.4 \% ($\eta_{T}$=42.0 \% and $\eta_{PV}$=18.4 \%).

During the simulation of the PV/T system, the sun light directly absorbed by the liquid was assumed to be constant across the liquid thickness. However, the calculations based on the Beer–Lambert law have shown, that most of the energy is absorbed in the upper stream layers of the liquid. Thus, the temperature of the upper layers of the liquid have to be considerably higher compared to the core of the stream. In order to estimate the impact of uneven light absorbtion by the liquid and its effect on the modelling by Eqs from \ref{EB} to \ref{q_loss2}, CFD (computational fluid dynamics) simulation in Ansys Fluent of the heat transfer processes in the PV/T system was performed at the following parameters:

\begin{enumerate}
  \item  tetralin solution (0.00299 kg kg$^{-1}$ of C$_{60}$) is a heat transfer fluid; 
  \item inlet temperature of fluid is 40 $^\circ$C and flow velocity is 0.03 m s$^{-1}$;
  \item heat generation by the PV cell was set as internal heat source. The value of the internal heat source was estimated at the average temperature of PV cell by equations from \ref{EB} to \ref{q_heat} and methods proposed by Otanicar et al. \cite{Otanicar2013} and An et al. \cite{an2017analysis};
  \item temperature dependent properties were used for the heat transfer calculation between liquid, glass cover and PV cell. The parameters listed in the Table \ref{Input data} were applied;
  \item energy absorbed by the liquid was set in following manner: case 1 --- internal heat source variable across the liquid thickness according to Beer–Lambert law (Equation \ref{I_trans}); case 2 --- constant internal heat source across the liquid thickness; case 3 --- internal heat source is absent (pure tetralin case).
\end{enumerate}

PV cell internal heat source for CFD simulation is expressed as following
 \begin{equation}\label{PV_IHS}
 p_{PV}=\frac{I_{trans} \cdot (1-\eta_{PV})}{\delta_{PV}}
\end{equation}
where, $\delta_{PV}$ is taken for CFD simulation thickness of the PV cell, m.

\begin{figure}
\centering
\center{\includegraphics[width=0.5\linewidth]{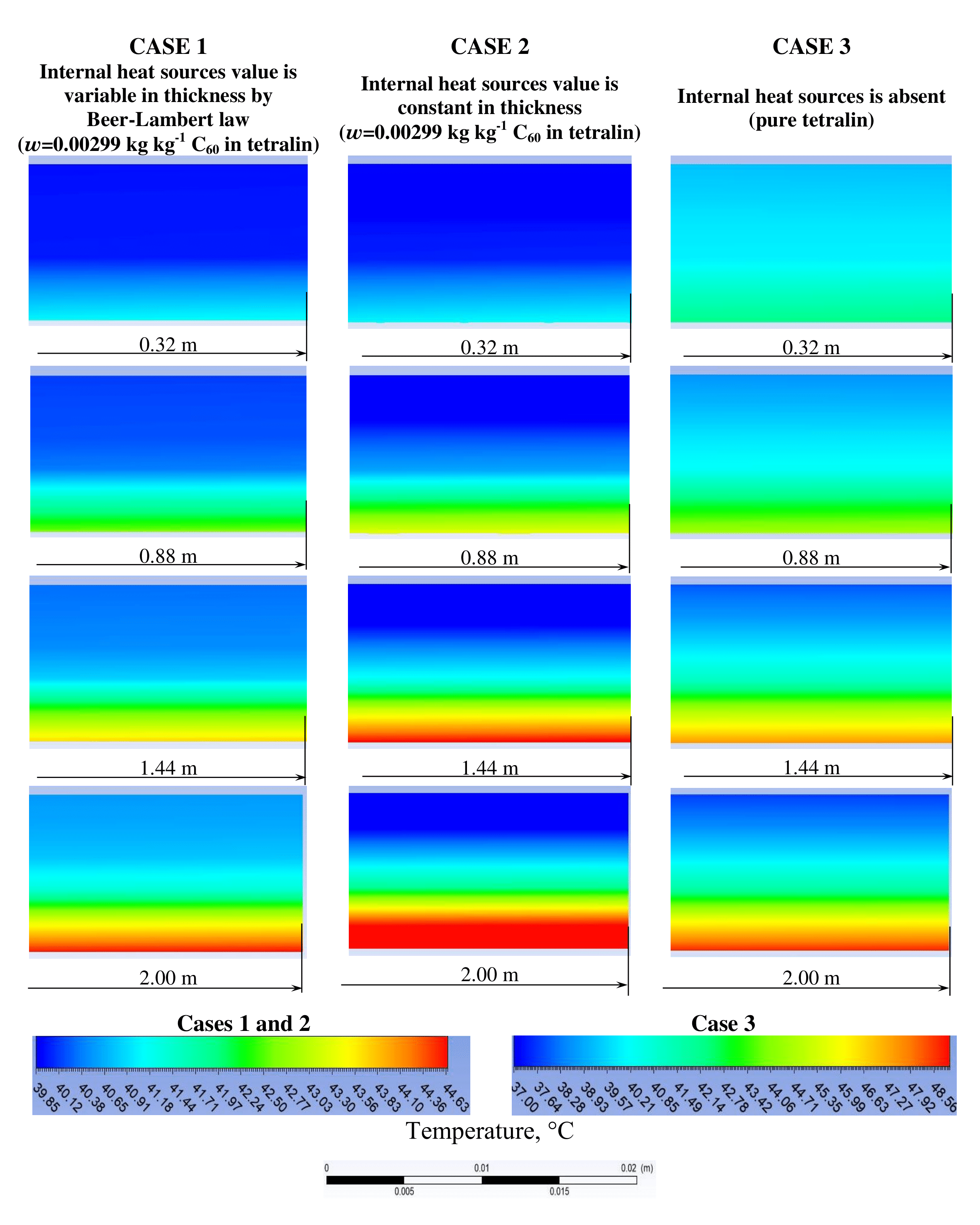}}
\caption{Results of CFD simulation for: vertical cross section of liquid channel along the flow direction at t$_{in}$=40 $^{\circ}$C and $v$=0.03 m s$^{-1}$}\label{CFD}
\end{figure}

Temperature fields of vertical cross sections along the length of PV/T system channel obtained during the CFD simulation for all cases of the internal heat sources from 1 to 3 are presented in Figure \ref{CFD}. Mean integral temperatures $\overline t_{liq}$, $t_{in}$, $t_{out}$, $\overline t_{PV}$ obtained through the CFD simulation were used in order to verify whether the type of internal heat source applied to the liquid significantly affect the results of calculation by the Eqs. \ref{EB} - \ref{q_heat}. PV cell output was again calculated using methods proposed by Otanicar et al. \cite{Otanicar2013} and An et al. \cite{an2017analysis} using obtained PV cell temperature by CFD simulation. The summary of the results obtained by CFD simulation and calculated by the Eqs. \ref{EB} to \ref{q_heat} are presented in Table \ref{Results}. As can be seen from Table \ref{Results} for the case 1, the most even distribution of the temperature was obtained. The temperature of PV cell for this case is the lowest and correspondingly  that provides certain benefits. However, the difference in the average PV cell temperature caused by the type of heat source applied to the liquid is not significant to change sufficiently the efficiency of the PV cell (see Figure \ref{Effic2}(a)). At the same time, average fluid temperature calculated by  Eqs. \ref{EB} to \ref{q_heat} perfectly match the CFD results.

\begin{table}
\caption{Results of the splitting PV/T system modelling and CFD simulation}
\label{tabular:timesandtenses}
\begin{center}
\begin{tabular}{llllll}
\hline
Parameters & modelling by Eqs & modelling by Eqs  & CFD simulation  & CFD simulation & CFD simulation \\ 
     & \ref{EB} - \ref{q_heat} Case 2 & \ref{EB} - \ref{q_heat} Case 3 & Case 1 &  Case 2   & Case 3 
\\ \hline
Average fluid temperature & 40.95 $^{\circ}$C & 40.82 $^{\circ}$C & 41.03 $^{\circ}$C & 41.02 $^{\circ}$C & 40.81 $^{\circ}$C \\
Average PV cell temperature & 47.3 $^{\circ}$C & 53.80 $^{\circ}$C & 45.02 $^{\circ}$C & 45.65 $^{\circ}$C & 49.88 $^{\circ}$C \\
Thermal output & 458.7 W m$^{-2}$ & 396.3 W m$^{-2}$ & 498.3 W m$^{-2}$ & 493.5 W m$^{-2}$ & 391.8 W m$^{-2}$ \\
PV cell output & 114.0 W m$^{-2}$ & 173.1 W m$^{-2}$ & 115.8 W m$^{-2}$ & 115.3 W m$^{-2}$ & 177.9 W m$^{-2}$ \\
$I_0 \cdot \eta_{opt}$ (energy balance check) & 711 W m$^{-2}$ & 711 W m$^{-2}$ & 754 W m$^{-2}$ & 742 W m$^{-2}$ & 700 W m$^{-2}$ \\
Energy balance deviation & - & - & 5.8 \%  & 4.4 \% & -1.5 \% \\
\hline
\end{tabular}
\end{center}
\label{Results}
\end{table}

From performed analyses is clearly seen that the way the light absorption was set (cases 1 to 3) do not influence significantly the results of modelling for the examined case using equations \ref{EB} - \ref{q_heat}. The maximum energy balance deviation do not exceed 5.8 \%. Nonetheless, for the other conditions, as well as for the fluids with higher and wider absorption spectrum this effect will be higher and have to be appropriately considered during modeling.

\section{Conclusions}
This study assesses the suitability of tetralin/C$_{60}$ solutions as a working fluid for hybrid photovoltaic-thermal solar energy harvesting. For this purpose, comprehensive analysis consisting of thermophysical and optical properties, single phase heat transfer, hydrodynamics and flat-plate PV/T collector efficiency was performed. The following conclusion can be formulated based on results obtained:

-The solutions of C$_{60}$ in tetralin have outstanding stability and simple synthesis method that is typical obstacle for most of the nanofluids proposed for thermal management applications;

-Absorption spectrum of the solutions is suitable for both direct solar absorption and especially for spectral splitting PV/T systems, due to the sharp spectrum cut off just before the ideal optical window of PV cells (around 650 nm);

-The viscosity of tetralin increases slightly due to the additives of C$_{60}$ while the thermal conductivity, density and specific heat capacity remains almost unchanged; 

-Energy feasibility of tetralin/C$_{60}$ solutions usage as the heat transfer fluid was analyzed using a new FoM, that consider pressure drop rise and heat transfer enhancement. It was found that at the low mass fractions of C$_{60}$ the heat transfer performance is insignificantly lower compared to pure tetralin. However, at the highest concentration of C$_{60}$, the heat transfer efficiency of tetralin/C$_{60}$ solutions is lower from the energy saving point of view; 

-Simulation of flat-plate PV/T collector revealed the total efficiency of 60.4 \% estimated with accordance to regulation (EU) No. 811/2013. It was also found that the tetralin/C$_{60}$ solution helps to balance solar energy transformation by the C$_{60}$ concentration proportionally between electrical and thermal energy output, when the total harvested energy remains nearly constant;

%====This efficiency was shown to be independent on C$_{60}$ concentration. It is also evident that market-available design of PV/T collector used for simulations does not realize the full potential of tetralin/C$_{60}$ nanofluid. Configurations with excess energy, such as concentrated PV, can be suggested as a better option.

-Life cycle analysis (LCA)  was performed in order to assessed tetralin/C$_{60}$ solutions in environmental perspective (see Appendix A) with estimation of environmental impacts and cumulative energy demand. It was found, that C$_{60}$ additives produce certain environmental costs. This is largely due to a non-optimal construction of PV/T collector, which was chosen based on market availability. Further researches should be focused on the PV/T collector design optimization as well as on the development of new ways of C$_{60}$ production that have lower environmental impact. This can help increasing attractiveness of tetralin/C$_{60}$ solutions for thermal management of flat-plate PV/T collectors from sustainability perspective.

\section*{CRediT authorship contribution statement}

Rita Adrião Lamosa \& Igor Motovoy: Investigation, Formal analysis, Visualization. Nikita Khliyev: Validation, Investigation, Formal analysis, Visualization. Artem Nikulin: Conceptualization, Writing original draft, Methodology, Formal analysis, Supervision, Investigation,
Visualization, Writing-review \& editing. Olga Khliyeva: Writing original draft, Methodology, Validation, Formal analysis, Investigation, Visualization, Writing-review \& editing. Ana S. Moita: Formal analysis, Supervision, Visualization, Writing-review\& editing. Janusz Krupanek: Formal analysis, Investigation, Visualization, Writing-review \& editing. Yaroslav Grosu: Formal analysis, Writing-review \& editing. Vitaly Zhelezny: Methodology, Formal analysis, Supervision, Writing-review \& editing. Antonio Luis Moreira: Supervision, Writing-review \& editing. Elena Palomo del Barrio: Supervision, Writing-review \& editing.

\section*{Declaration of Competing Interest}
The authors declare that they have no known competing financial interests or personal relationships that could have appeared to influence the work reported in this paper.

\section*{Acknowledgements}
Authors would like to acknowledge the following projects for the financial support of this study:

-Fundacao para a Ciencia e a Tecnologia (FCT) projects UID/EEA/50009/2019 and PTDC/EME-SIS/30171/2017;

-National Research Foundation of Ukraine, project No. 2020.02/0125;

-FSWEET-TES project (RTI2018-099557-B-C21), funded by FEDER/Ministerio de Ciencia e Innovación – Agencia Estatal de Investigación;

-Elkartek CICe2020 project (KK-2020/00078) funded by Basque Government.

Authors thank goes to professor Fernanda Margarido (IN+, IST, Portugal) for providing the facilities of the laboratory of Industrial Ecology and Sustainability and Sandra Dias (IN+, IST, Portugal) for her help with thermal conductivity cell oxidation.

\newpage
\appendix
\large{\textbf{Appendix}}
\setcounter{table}{0}
\setcounter{figure}{0}

\section{Life cycle analysis}
\renewcommand{\thetable}{\Alph{section}\arabic{table}}
\renewcommand{\thefigure}{\Alph{section}\arabic{figure}}

To screen the environmental performance of the studied solution a Life Cycle Analysis (LCA) was performed for a simplified system of nanofluid application for combined photovoltaic and solar panel. Life Cycle Analysis \cite{ISO14044}, is a technique to assess environmental impacts associated with all the stages of the life-cycle of a commercial product, process, or service through the product's manufacture, distribution and use, to the recycling or final disposal of the materials. LCA is performed in the following steps:  \textit{(i)} technological system definition; \textit{(ii)} data collection; \textit{(iii)} risk evaluation (risk characterization) and impacts quantification (Life Cycle Impact Assessment—LCIA), \textit{(iv)} results interpretation. The system assessed in this work is limited to production of tetralin/C$_{60}$ solutions. The other stages such as application in photovoltaic panel, service life, dismantling of the functional elements after the service life and management of wastes was assumed as comparable for all variants. LCA was performed with regard to ISO Standard 14040:2009 using SimaPro tool and EcoInvent 3.0 database.

Solutions in this work are developed for the benefit of application for combined photovoltaic and solar panel. The Life Cycle Assessment is performed with regard to the results of the research done in this study.

\subsection{Goal and scope of the life cycle analysis}

The goal of the study is to assess the environmental impacts and compare use of tetralin/C$_{60}$ solutions application. The following options (weight concentrations) are considered: 

\begin{enumerate}
  \item  pure tetralin;
  \item tetralin/C$_{60}$ solutions containing  0.0005, 0.001 and 0.00299 \textbf{\% of C$_{60}$}. 
  \end{enumerate}

Functional unit is defined as 1 kg of working fluid in the integrated photovoltaic and solar panel. The assessment is performed from cradle to gate. The environmental effect (benefit) of application of the solution application was not determined in the study. The working time life of the solutions is also not determined. Considering these aspects the only one taken into account is the phase of production of tetralin and production of fulleren C$_{60}$. The post-use phase was assumed as similar for all variants. Although, in the study, there is assessed potential application in existing or future energy production systems, and  uncertainties had to be taken into account, but for the purpose of the study the best case scenario for C$_{60}$ production via tetralin pyrolysis was taken into account. According to literature \cite{anctil2012life} this way of production is characterized with the best environmental performance in terms of Cumulative Energy Demand and can be viewed as a benchmark for such a system. 

\subsection{Data collection} 
Data for the LCA study have been effectively collected in agreement with the requirements of the method and the definition of the system boundaries. To characterize the process the data in the available literature was used. For tetralin production the pathway of naphthalene dehydrogenation is considered as the main process \cite{alford2008fullerene} and tetralin pyrolysis for production of C$_{60}$ fullerene \cite{anctil2011material}. Inventory data for the study are presented in the Table \ref{tabular:Data inventory}.

\begin{table} 
\caption{Data inventory for the scenarios with solutions application calculated per kg of functional fluid prepared based on literature \cite{alford2008fullerene,anctil2011material}}
\label{tabular:Data inventory}
\begin{tabular}{|l|l|l|l|l|l|}
\hline
\multirow{2}{*}{Process}                   & \multicolumn{1}{c|}{\multirow{2}{*}{Material/Energy}} & \multicolumn{1}{c|}{\multirow{2}{*}{Unit}} & \multicolumn{2}{c|}{Scenarios}         & \multicolumn{1}{c|}{\multirow{2}{*}{Comment}} \\ \cline{4-5}
                                           & \multicolumn{1}{c|}{}                                 & \multicolumn{1}{c|}{}                      & tetralin & tetralin/C$_{60}$ variants range & \multicolumn{1}{c|}{}                         \\ \hline
\multirow{4}{*}{Tetralin production}       & Tar                                                   & kg                                         & 2.54     & 2.62-3.07*                   & Ecoinvent 3                                     \\ \cline{2-6} 
                                           & Heat                                                  & kWh                                        & 0.63     & 0.66-0.77*                   & Ecoinvent 3                                     \\ \cline{2-6} 
                                           & Steam                                                 & kg                                        & 0.16     & 0.16-0.20                   & Ecoinvent 3                                     \\ \cline{2-6} 
                                           & Electricity                                           & kWh                                        & 0.19     & 0.19-0.23                   & ELCD                                  \\ \hline
\multirow{7}{*}{C$_{60}$ fullerene production} & Oxygen                                                & kg                                         & ---      & 0.03-0.20               & EcoInvent                                     \\ \cline{2-6} 
                                           & Electricity                                           & kWh                                        & ---      & 0.08-0.45                   & ELCD                                     \\ \cline{2-6} 
                                           & Steam                                                 & kg                                        & ---      & 0.12-0.73                   & Ecoinvent 3                                     \\ \cline{2-6} 
                                           & o-xylene                                              & kg                                         & ---      & 0.01-0.06                   & Ecoinvent 3                                     \\ \cline{2-6} 
                                           & Rail transport                                        & tkm                                        & ---      & 0.001-0.004                 & Ecoinvent 3                                     \\ \cline{2-6} 
                                           & Road transport                                        & tkm                                        & ---      & 0.003-0.019                 & Ecoinvent 3                                     \\ \cline{2-6} 
                                           & Hydrocarbons emissions               & kg                                         & ---     &  0.038-0.223                   &      Ecoinvent 3**                                         \\ 
                                           &expressed as C                &                                        &     &                     &                                            \\ \cline{1-6}

                                           & Hazardous wastes                                      & kg                                         & ---     &  0.002-0.012                   &      Ecoinvent 3                                         \\ \hline                                           
\end{tabular}

\raggedright{* economic allocation}

\raggedright{** modified data for carbon black production}
\end{table}

\subsection{Risk evaluation and impacts quantification} 
\begin{figure}
\centering
\center{\includegraphics[width=0.7\linewidth]{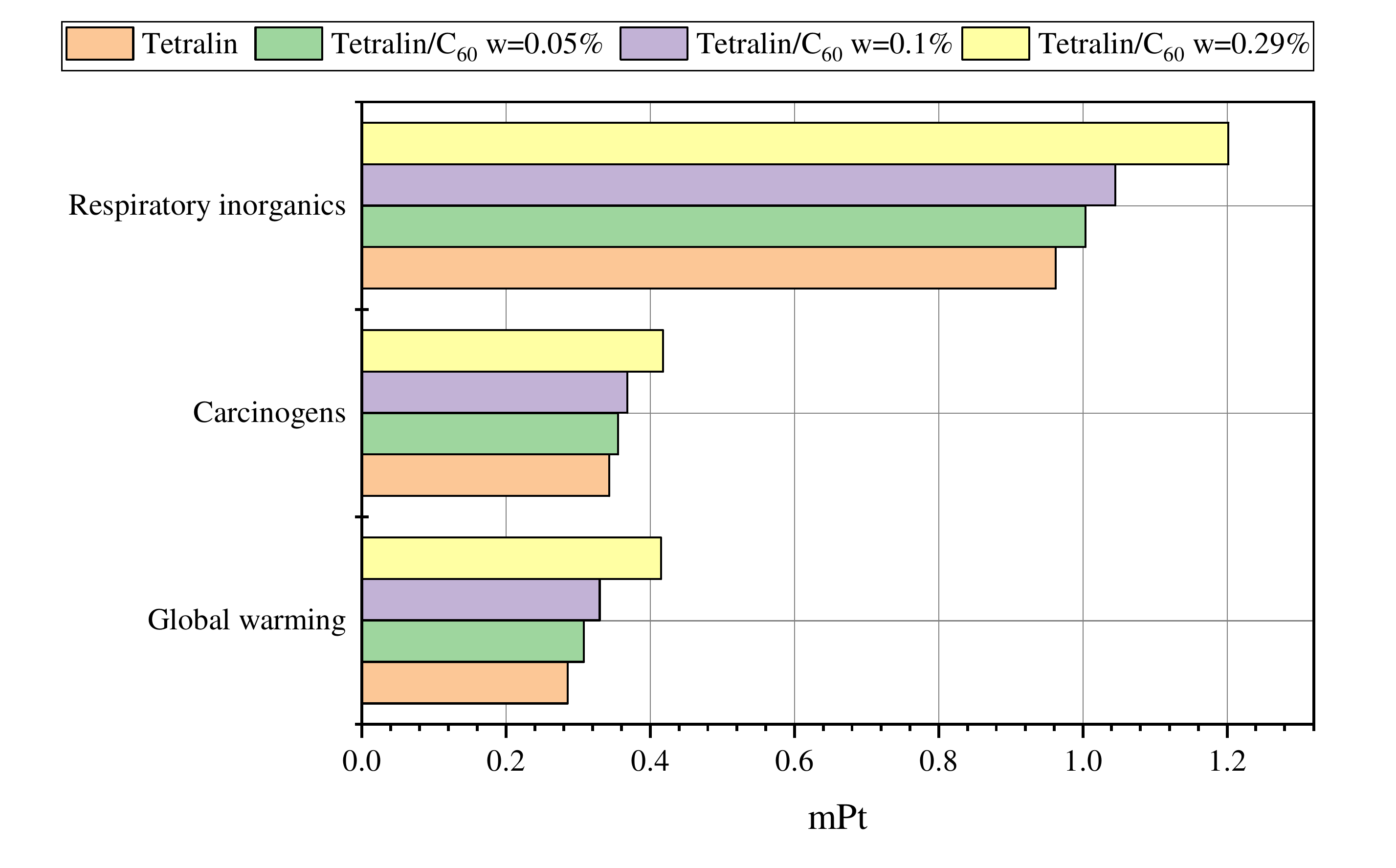}}
\caption{Characterisation of the main impacts for compared scenarios (mPt – damage standardized measure)} \label{Damage}
\end{figure}

The characterization method applied in this work to evaluate the environmental impacts is IMPACT 2002+  and Cumulative Energy Demand. The results are presented in the Figure \ref{Damage}. In general the tetralin/C$_{60}$ variants are characterized with higher environmental impacts. Main impacts are related predominantly to processing of carbon materials in tetralin production and to energy use during production of tetralin and C$_{60}$.

It should be noted that due the lack of data, the LCA does not consider specific aspects of nanoparticles toxicity to humans and the environment. Risk assessment studies carried out for fullerene show that this aspect has to be taken into account \cite{gerbinet2014life} and appropriate risk management measures applied during production, use and waste disposal phases \cite{van2012stoffenmanager}. 

\begin{figure}
\centering
\center{\includegraphics[width=0.7\linewidth]{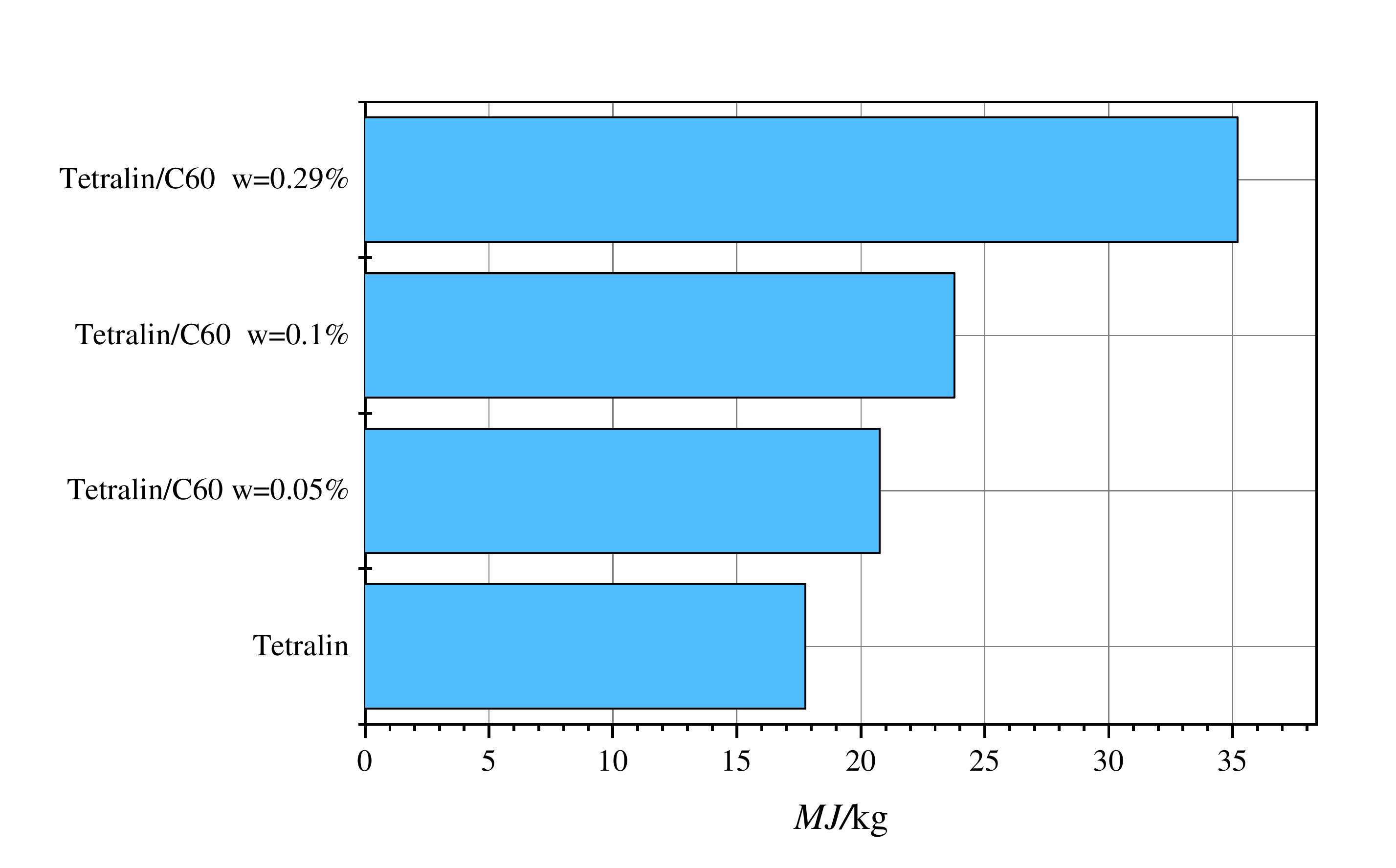}}
\caption{Characterisation of impact for Cumulative Energy Demand indicator} \label{Cumulative}
\end{figure}

The similar pattern is shown for Cumulative Energy Demand (Figure \ref{Cumulative}). 

\subsection{Interpretation} 

The characterization method applied in this work to evaluate the environmental impacts is IMPACT 2002+ and cumulative energy demand. The results are presented in the Figure \ref{Damage}. In general the tetralin/C$_{60}$ variants are characterized with higher environmental impacts. Main impacts are related predominantly to processing of carbon materials in tetralin production and to energy use during production of tetralin and C$_{60}$.

The results of LCA study based on literature data (see Table \ref{tabular:Data inventory}) are of indicative character. They indicate the reasonability from sustainability perspective of application of the studied solutions. The main conclusion from the LCA study is that there is the environmental impact rise related to production phase versus the concentration of C$_{60}$ in tetralin. The main environmental impacts related to use of the tetralin/C$_{60}$ solutions are: respiratory inorganics, carcinogens and global warming potential. From environmental point of view there is no strong indication to recommend the use of fullerene C$_{60}$ as additive to tetralin to be used for the specific case studied here. The reason is that the potential environmental benefits of nanoparticle application were not determined as sufficient to balance the environmental costs in the Life Cycle Perspective. The benefits from using the C$_{60}$ in the specific case of PV/T system considered here is very low in comparison with pure tetralin. In recent studies \cite{anctil2016material} it is underlined that the embodied energy of all fullerenes are an order of magnitude higher than most common chemicals and, therefore, are likely to influence the embodied energy of the product they will be used in, even though they might only represent a small fraction of its total mass. The benefits from using the C$_{60}$ in the specific case of PV/T system considered here is very low in comparison with pure tetralin.

%The case study simulation has shown that the tetralin/C$_{60}$ solution helps to balance solar energy transformation by the C$_{60}$ concentration proportionally between electrical and thermal energy output. The lack of clear functional benefits related to energy efficiency in specific applications is the main drawback in the assessment. 

%On the other hand the LCA results does not rule out the reasonability to use tetralin/C$_{60}$  solutions when the specific characteristics can be efficiently utilized in concrete applications. Theoretically, considering only the one factor of environmental production costs of nanofluid it is of minor importance. To have an environmental payback in the life time of PV/T system for the service life the gain in higher energy performance should be at least 5-15 MJ. Assuming for example ratio between fluid volume and exergy output 1:5000 during a year of the PV/T system and 1 \% the potential benefits can be very low to justify use of tetralin/C$_{60}$ solution. Higher efficiency rate for low C$_{60}$ concentrations applied in PV/T systems can prove to be beneficial or at least neutral from environmental point of view. 

It has to be underlined that apart of the environmental impact during production phase there are other essential environmental impacts related to use and the after service life phase. Moreover, the waste management of the mixture after service life has to be considered. Taking into account the overall assessment further possibilities of nanomaterials application should be studied considering novel opportunities \cite{diwania2020photovoltaic}.

%Considering the use phase there is expected increase of the viscosity and pressure loses as well as lower heat transfer efficiency of the solutions with higher content of C60. Here are also important factors related to life time of application of the mixture (its working stability) the specific environmental gains related to energy efficiency for consumers and working conditions.

%Regarding this crucial should be further analysis performed for the use phase in a given PV/T system. There are, moreover uncertainties related to health and ecological risks and the data used in the assessment. 

\afterpage{
\bibliographystyle{elsarticle-num-names} %In sequence

\bibliography{references}
}

\end{document}